\documentstyle[pra,aps,eqsecnum,amsmath,floats,12pt]{revtex}
\tighten
 \input BoxedEPS
\SetRokickiEPSFSpecial  %% for dvips by Tom Rokicki, for VMS
\HideDisplacementBoxes  %% hides the rectangular figure outlines

\def\Fig#1#2#3#4{\begin{figure}
                        $$\BoxedEPSF{#1 scaled #2}$$
                        \caption{#4}
                        \label{fig:#3}
                        \end{figure}}

\newcommand\Figcite[1]{Fig.\thinspace\ref{fig:#1}}
\def\cal#1{{\mathcal{#1}}}

\def\rdr{\hat\rho^\dagger_i\hat\rho^{\phantom{\dagger}}_{i+1}}
\def\rrd{\hat\rho^{\phantom{\dagger}}_i\hat\rho^\dagger_{i+1}}

\begin{document}
%\draft

\title{Quantum Computation and Decision Trees}

\author{Edward Farhi\thanks{This work was supported in
part by The Department of Energy under cooperative agreement
DE-FC02-94ER40818.}}
\address{Center for Theoretical Physics\\
Massachusetts Institute of Technology\\
Cambridge, MA  02139\\
{\small\tt farhi@mitlns.mit.edu}}
\author{Sam Gutmann}
\address{Department of Mathematics\\
Northeastern University\\ 
Boston, MA 02115\\
{\small\tt sgutm@nuhub.neu.edu}}
\date{July 1997; revised March 1998}
\maketitle

\begin{abstract}
Many interesting computational problems can be reformulated in terms of
decision trees.  A natural classical algorithm is to then run
a random walk on the tree, starting at the root,  to see if the tree contains a node~$n$
levels from the root.  We devise a quantum mechanical algorithm that
evolves a state, initially localized at the root, through the tree.  We prove that if the
classical strategy succeeds in reaching level $n$ in time polynomial in $n$,
then so does the quantum algorithm. Moreover, we find examples of trees for which
the classical algorithm requires time exponential in
$n$, but for which the quantum algorithm succeeds in polynomial time.  The examples
we have so far, however, could also be solved in polynomial time by different classical
algorithms.
\medskip

\centerline{MIT-CTP-2651,~~quant-ph/9706062}

\end{abstract}


\section{Introduction}
Many of the problems of interest to computation experts are, or are reducible to,
decision problems.  These are problems that for a given input require the
determination of a yes or no answer to a specified question about the input.  For
example the traveling salesman problem is (polynomial time) equivalent to the
decision problem that asks whether or not for a given set of intercity
distances there is a route passing through all of the cities whose length is
less than a given fixed length.  Another example that we will later use for
concreteness in this paper is the $0-1$ integer programming problem called
``exact cover"\cite{ref:1}.  Here we are given an $m$ by $n$ matrix, $A$, all of
whose entries are either 0 or 1.  The number of columns $m$ is $\le n$.  We are
asked if there exists a solution to the $m$ equations
\begin{equation}
\sum_{k=1}^n A_{jk} x_k =1 \qquad \mbox{for~}j=1,m
\label{eq:1}
\end{equation}
with the $x_k$ restricted to be 0 or 1.  The brute force approach to this
problem is to try the $2^n$ possible choices of $\vec{x}=(x_1,\dotsc, x_n)$.  For
each choice of
$\vec{x}$, checking to see if Eq.~(\ref{eq:1}) is satisfied takes at most of
order $mn$ operations, which is polynomial in the input size.  However, checking
all $2^n$ possible choices for $\vec{x}$ is prohibitively time consuming even
for moderately large values of $n$.

For the exact cover problem, with a given instance of the input matrix $A$, it
is not actually necessary to check all $2^n$ values of $\vec{x}$ to see if
Eq.~(\ref{eq:1}) can be satisfied.  Note that generically $x_1$ can take the
values 0 or 1 and $(x_1, x_2)$ can have the values $(0,0), (0,1), (1,0)$ or
$(1,1)$.  However, suppose that for some $j$ the matrix $A$ has $A_{j1} =
A_{j2} =1$.  In this case the choice $x_1 = x_2 =1$ is eliminated and no
 $\vec{x}$ of the form $(1,1,x_3,\dotsc, x_n)$ need be tried.  If we
consider  $\vec{x}$'s  that begin with $x_1,x_2,\dotsc, x_\ell$ then if for
some $j$ we have $\sum^\ell_{k=1} A_{jk} x_k \ge 2$, then any  $\vec{x}$ 
beginning with $x_1,x_2,\dotsc, x_\ell$ is eliminated.  We can picture this in
terms of a decision tree as follows.
\Fig{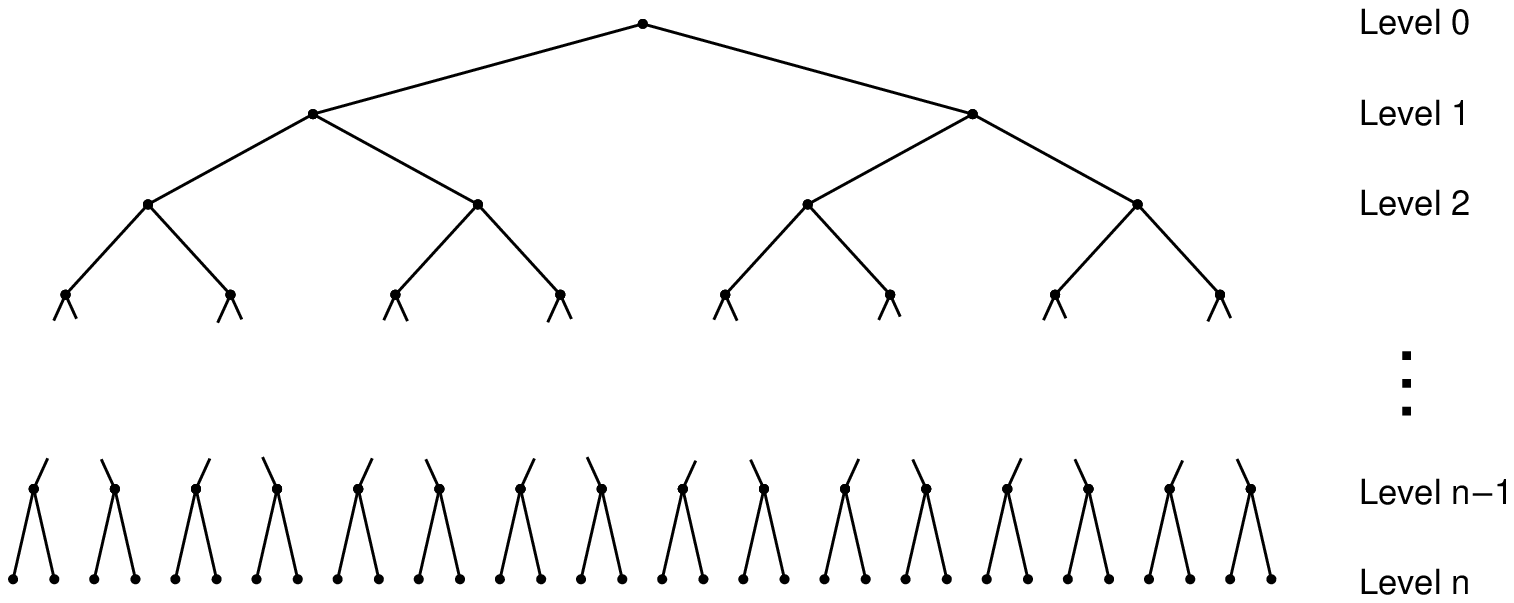}{900}{Figi}{The underlying branching tree. At level $m$ there are
$2^m$ nodes.}%
  Before imposing any constraints we
construct an underlying branching tree.  This tree starts at the top with one
starting node that branches to two nodes corresponding to the two choices for
$x_1$.  This then branches to the four choices for $(x_1, x_2)$ and so on until
we have all $2^n$ choices for $(x_1 \cdots x_n)$ at the $n^{\rm th}$ level.
However if we impose the constraints and see that a particular node is
eliminated, then we can also eliminate all nodes connected to that node that
lie below it in the tree.  The decision tree is the underlying branching tree
that has been trimmed as a result of the constraints.
Note that the exact cover problem has a solution if and only if the
decision tree has one or more nodes left at the bottom ($n^{\rm th}$) level.

\Fig{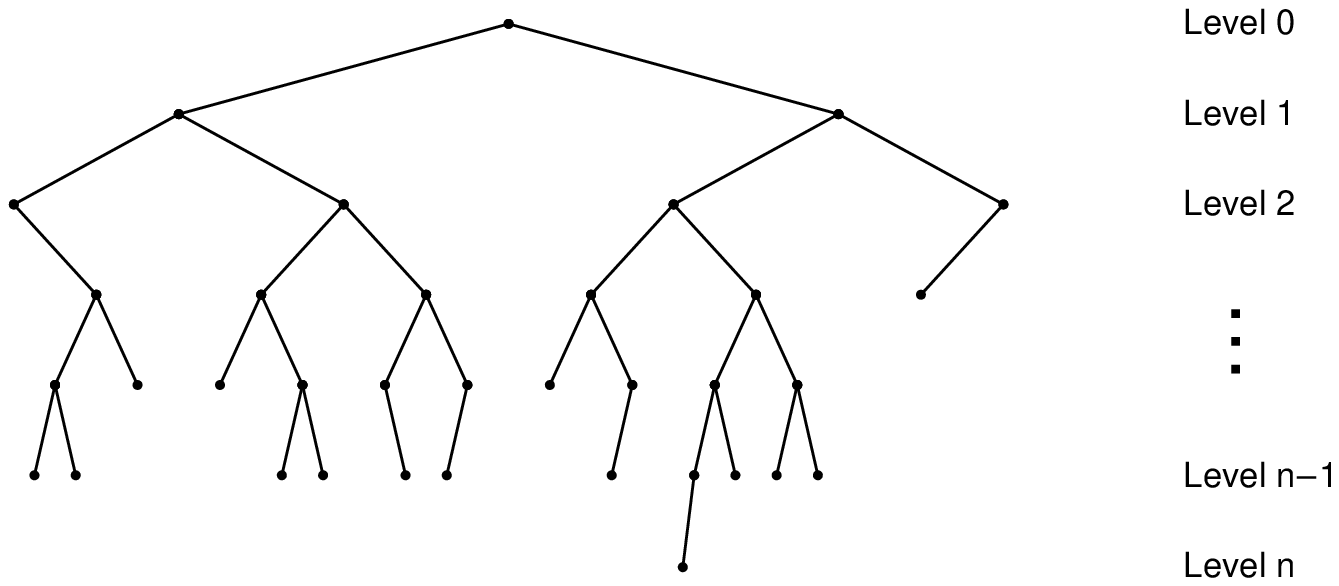}{900}{Figii}{An example of a decision tree, $T_n$, with one node at
level~$n$. For aesthetic reasons we will no longer put breaks in trees -- they are still
to be thought of as being many levels deep.}%

More generally we view decision problems as having an underlying bifurcating
branching tree with $n$ levels as in Fig.\thinspace\ref{fig:Figi}.  The specific form
(or instance) of the problem imposes constraints that eliminates  nodes from the tree
as in \Figcite{Figii}.  When a node is excluded the whole
branch with that node as its topmost point is also cut from the tree.  The decision
question we wish to answer is ``are there any nodes left at the
$n^{\rm th}$ level after all constraints have been imposed?''

Consider a family of decision problems indexed by a size $n$. Particular
instances of the problem of size $n$ give rise to particular decision trees
that either have or do not have nodes at the
$n^{\rm th}$ level.  The computational concern is how much time, or how many
algorithmic steps, are required to answer the decision question as $n$ gets big. 
Roughly speaking if the time grows like
$n^A$ for fixed $A > 0$, the problem is considered easy whereas if the time
grows like
$a^n$ with $a >1$, the problem requires an ``exponential amount of time'' and is
considered computationally hard.

One approach to solving a decision problem is to systematically check every path
that starts at the top of the tree and moves downward through the tree.  If a
path reaches a dead end you try the next path (in some list of paths) until you
find a path that has a node at the $n^{\rm th}$ level or else, after having
checked all paths, you discover that the answer to the decision question is
``no".  An alternative to systematically exploring the whole tree is to move
through the tree with a probabilistic rule.  For example you could use the rule
that if you are at a given node you move to the other nodes that are connected
to it with equal probability.  Thus if you are at a node that connects to two
nodes below it you have a 1/3 chance of moving back up the tree; if the node
connects to just one below you have a 1/2 chance of moving back up whereas if
the node is a dead end you definitely move back.  If you start at the top of the
tree and move with this probabilistic rule you will eventually visit every node
in the tree. 

Consider a family of decision trees that are associated with underlying
branching trees that are $n$ levels deep.  An individual instance of the
decision tree either has or does not have nodes at the $n^{\rm th}$ level.  If
it does have nodes at the $n^{\rm th}$ level and we use a probabilistic rule for
moving through the tree, then we say that the tree is penetrable if there is a
good chance of reaching the $n^{\rm th}$ level in not too great a time. More
precisely we define the family of trees as penetrable if: 
$$\begin{minipage}{.75\hsize}{There exist $A,B > 0$ such that for those trees with a
node (or nodes) at the 
$n^{\rm th}$ level  there is a 
$t < n^A$ with the probability of being at the $n^{\rm th}$
level by $t$  greater than 
  $(1/n)^B$.}
\end{minipage} \eqno{(\rm P)} $$
This means that in polynomial time the probability of reaching the $n^{\rm th}$
level is at worst polynomially small. 
 If (P) is met, then by running the process order $n^B$ times we achieve a probability
of order~1 of reaching the $n^{\rm th}$ level in time $n^{A+B}$. 
 If (P) is not met it means that  it
either takes more than polynomial time to reach the $n^{\rm th}$ level or the
probability of reaching the
$n^{\rm th}$ level is always smaller than $(\frac{1}{n})$ to any power. 
Therefore if condition (P) is not met instances of the trees with nodes at the
$n^{\rm th}$ level cannot practically be distinguished from instances with no
nodes at the $n^{\rm th}$ level.  
In this case the corresponding decision problem is not solvable in polynomial time by
this algorithm. 
We will divide families of decision trees indexed by
$n$ into two classes, those that satisfy (P) and those that do not, which we call
impenetrable.

We are interested in using quantum mechanics to move through decision trees.  We
imagine that  nodes of the decision tree correspond  to  quantum states, which 
give a basis for the Hilbert space.  We further imagine constructing a
Hamiltonian
$\hat{H}$ with nonzero off-diagonal matrix elements only between states that are
connected in the corresponding decision tree.  (We will be more specific about
constructing the Hilbert space and $\hat{H}$ later.)  We start the quantum system
in the state corresponding to the topmost node and let it evolve with its time
evolution determined by $\hat{H}$ so that the unitary time evolution operator~is
\begin{equation}
\hat{U}(t)=\exp(-i\hat{H} t) \ \ .
\label{eq:1.2}
\end{equation}
At any time $t$ we have a pure state that can be expressed as a (complex)
superposition of basis states corresponding to the nodes.  Given $\hat{H}$ and the
initial state, the probability (the amplitude squared) of finding the system at the
$n^{th}$ level at time $t$ is determined.  We then say that a family of trees
indexed by size $n$ is quantum penetrable if condition $(P)$ is met and it is
quantum impenetrable if condition $(P)$ is not met.

In the next section we will give a specific form for the quantum Hamiltonian
$\hat{H}$ and then prove that any family of trees that is classically
penetrable is associated with a closely related family of trees that is quantum
penetrable.  This will demonstrate that our model for quantum mechanically
solving decision problems is at least as powerful as the classical probabilistic
method.  In Section Three we will go further and give an example of a family of
decision trees that is classically impenetrable but which is quantum
mechanically penetrable.  This means that the quantum penetration is
exponentially faster than the corresponding classical penetration for these
trees.  However, we have not yet identified general characteristics of a problem that
guarantee that its associated decision trees are quantum penetrable. Furthermore, for
the example considered, the problem associated with the classically impenetrable trees
can be reformulated so that it is computationally simple to solve by an alternative,
classical method.

In Section Four we discuss the construction of the Hilbert space and the
Hamiltonian $\hat{H}$.  The usual paradigm for quantum computation\cite{ref:2}
envisages a string of, say, $\ell$ spin-1/2 particles that gives rise to a
$2^\ell$-dimensional Hilbert space.  Each elementary operation is a unitary
transformation that acts on one or two spins at a time.  We will show that the
Hilbert space for our system can be constructed using $\ell$ spin 1/2 particles 
just as in a conventional quantum computer.  Furthermore, for  a large class of
problems, the Hamiltonian that we construct is a sum of Hamiltonians that act on a
fixed number of spins.  In this sense\cite{ref:3} our quantum evolution through
decision trees lies in the framework of conventional quantum computation.
\bigskip\bigskip

\section{Classical vs.\ Quantum Evolution through Trees}

In the introduction we discussed a classical (that is, non-quantum) probabilistic
rule for moving through decision trees.  Here we are going to be more specific
and state the rule in a way that gives rise to a continuous time Markov
process.  The rule is simply that if you are at a given node then you move to a
connected node with a probability per unit time
$\gamma$ where $\gamma$ is a fixed, time independent, constant.  This means that
in a time $\epsilon$ where $\gamma\epsilon \ll 1$, the probability of moving to
a connected node is $\approx \gamma\epsilon$.  Using a continuous time process, as
opposed to saying that you move at discrete times, will make it easier when we
compare with the continuous time evolution dictated by the unitary operator in
(\ref{eq:1.2}).

We are now going to introduce some formalism that looks quantum mechanical but
we are going to apply it to describe the classical Markov process.  Suppose we
are given a decision tree that has $N$ nodes.  ($N$ may be as large as
$2^{n+1}$ where $n$ is the number of levels.)  Index the nodes in some way by
the integers $a=1,\dotsc,N$.  Corresponding to the tree we construct an
$N$-dimensional Hilbert space that has an orthonormal basis $\{|a\rangle\}$
with $a=1,\dotsc,N$ and accordingly $\langle a|b\rangle=\delta_{ab}$.  Now we
define a Hamiltonian $\hat{H}$ through its matrix elements in this basis:
\begin{eqnarray}
\langle b|\hat{H}|a\rangle & = & \left\{ \begin{array}{rl} -\gamma & \mbox{\quad
for
$a\neq b$ if node
$a$ is connected to node $b$} \\ 
0 & \mbox{\quad for $a\neq b$ if node $a$ is not
connected to node $b$} \end{array} \right. \nonumber \\
\langle a|\hat{H}|a\rangle & = &  \left\{ \begin{array}{rll} 3\gamma &
\mbox{\quad node $a$ is connected to three other nodes} \\ 2\gamma & \mbox{\quad
node $a$ is connected to two other nodes} \\
 \gamma & \mbox{\quad node $a$ is
connected to one other node\ \ .}
\end{array} \right. 
\label{eq:3}
\end{eqnarray}

Return to the classical probabilistic rule for moving through a fixed tree and
let
\begin{equation}
p_{ba}(t) = \mbox{Prob (go from $a$ to $b$ in time $t$)} \ \ .
\label{eq:2.2}
\end{equation}
For a time $\epsilon$ where $\gamma\epsilon \ll 1$ we have
\begin{equation} p_{ba}(\epsilon)  =  \left\{ \begin{array}{ll} -\epsilon
\langle b|\hat{H}|a\rangle + {\cal O}(\epsilon^2) & \mbox{\quad for $b\neq a$} \\
1 - \epsilon \langle a|\hat{H}|a\rangle +{\cal O}(\epsilon^2) & \mbox{\quad for
$b=a$}
\end{array}
\right.
\label{eq:2.3}
\end{equation}
as a consequence of the definition of $\hat{H}$.  For a classical Markov
process, the probability of moving depends only on current position, not on
history, so we have for any $t_1$ and $t_2$,
\begin{equation}
p_{ba} (t_1+t_2)=\sum_c p_{bc}(t_2)p_{ca}(t_1) \ \ .
\label{eq:2.4}
\end{equation}
Therefore
\begin{equation}
p_{ba}(t+\epsilon)=\sum_c p_{bc}(\epsilon) p_{ca}(t)
\label{eq:2.5}
\end{equation}
which for $\epsilon$ small gives
\begin{equation}
p_{ba}(t+\epsilon)=p_{ba}(t)-\epsilon \sum_c \langle b|\hat{H}|c\rangle
p_{ca}(t)+{\cal O}(\epsilon^2)
\label{eq:2.6}
\end{equation}
where we have used (\ref{eq:2.3}).  We see therefore that $p_{ba}(t)$ obeys the
differential equation
\begin{equation}
\frac{d}{dt} p_{ba}(t)=-\sum_c\langle b|\hat{H}|c\rangle p_{ca}(t)
\label{eq:2.7}
\end{equation}
with the boundary condition $p_{ba}(0)=\delta_{ab}$.  The solution to
(\ref{eq:2.7}) is
\begin{equation}
p_{ba}(t)=\langle b|e^{-\hat{H} t}|a\rangle \ \ .
\label{eq:2.8}
\end{equation}

Again, $p_{ba}(t)$ given by (\ref{eq:2.8}) is the {\bf classical} probability of
going from $a$ to
$b$ in time $t$ if you move through the tree with a probability per unit time
$\gamma$ of moving to a connecting node.  As a check we should have that
\begin{equation}
\sum_b p_{ba}(t)=1 \ \ .
\label{eq:2.9}
\end{equation}
To see that this is the case note that $\hat{H}$ defined by (\ref{eq:3}) has a
zero eigenvector,
\begin{equation}
|E=0\rangle=\frac{1}{\sqrt{N}} \sum^N_{b=1} \ |b\rangle \ \ .
\label{eq:2.10}
\end{equation}
Therefore
\begin{eqnarray}
\sum_b p_{ba}(t) & = & \sqrt{N} \; \langle E=0|e^{-\hat{H}t}|a\rangle \nonumber
\\ & = & \sqrt{N} \; \langle E=0|a\rangle \label{eq:2.11} \\[1ex] 
& = & 1 \ \ . \nonumber
\end{eqnarray}

We have constructed the Hamiltonian $\hat{H}$ because of its utility in
describing a classical Markov process.  We now propose using the same
Hamiltonian $\hat{H}$ to quantum mechanically evolve through the tree.  Let
\begin{equation}
A_{ba}(t)=\langle b|e^{-i\hat{H}t}|a\rangle
\label{eq:2.12}
\end{equation}
be the quantum amplitude to be found at node $b$ at time $t$ given that you are 
at node $a$ at time $0$.  In this case the probability is $|A_{ba}(t)|^2$ with
\begin{equation}
\sum_b|A_{ba}(t)|^2=1
\label{eq:2.13}
\end{equation}
as a consequence of the fact that $\hat{H}$ is Hermitian.  With this quantum
Hamiltonian we will now show that if a family of trees is classically penetrable
then there is a related family of trees that is also quantum mechanically penetrable.   

Imagine we are given a family of decision trees $\{T_n\}$ where each $T_n$ is $n$
levels deep and does have nodes at the $n$-th level.  For simplicity we will
take the worst case possible and assume that there is only one node at level
$n$.  In order to establish our result we are going to consider another family
of  trees $\{T'_n\}$ where each $T'_n$ is obtained from $T_n$ by
appending a semi-infinite line of nodes to the starting node of $T_n$.
\Fig{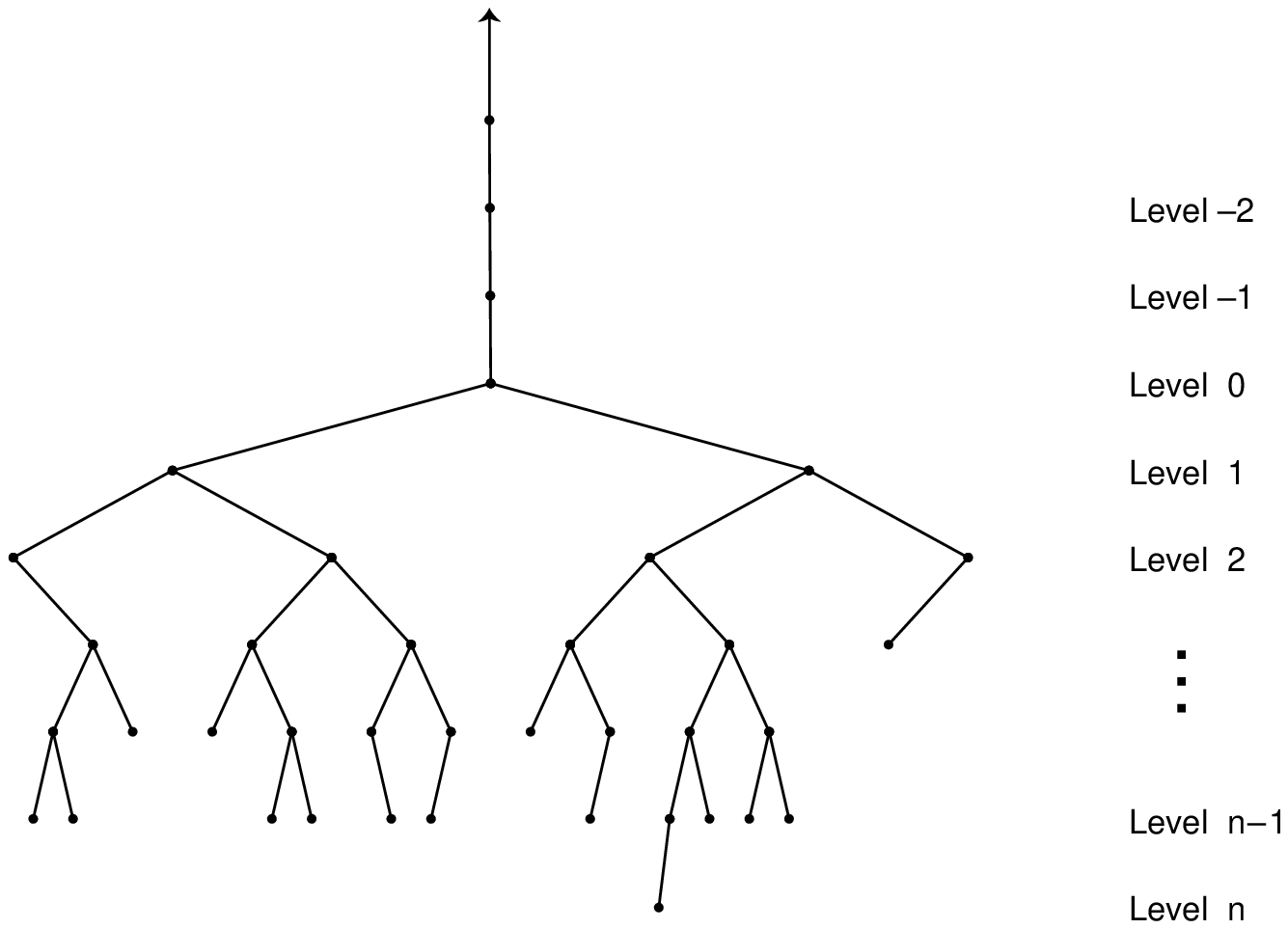}{900}{Figiii}{The tree $T'_n$ obtained from the tree $T_n$ of
Fig.\thinspace2 by appending a semi-infinite line of nodes at the starting node
of~$T_n$.}%
 The rule for classically moving on the semi-infinite line is the same
as the rule for moving on the rest of the tree: with a probability per unit time
$\gamma$ you move to an adjoining node. 

We can see that if $\{T_n\}$ is classically penetrable so is $\{T'_n\}$. 
Roughly speaking, starting at $0$ on $T'_n$, the probability of reaching the
$n^{th}$ level is not appreciably reduced because of the time some paths spend
on the semi-infinite line.  (We now prove this statement, but the reader who is
already convinced that it is true can skip to the next paragraph.) Suppose that
for
$\{T_n\}$ we have condition (P) so that
\begin{equation}
\mbox{Prob (go from $0$ to $n$ in time $t$)} \geq \frac{1}{n^B}
\label{eq:16new}
\end{equation}
for some $\gamma t\leq n^A$.  At level 1 of the decision tree only one of the two
nodes is on the branch that contains $n$, the unique node at level $n$.   
Denote this level 1 node by $\bar{1}$.  Now for each path (on $T_n)$ that
reaches
$n$ from 
$0$ in time $t$ there is a time $t-s$ at which the path last jumps from $0$ to
$\bar{1}$.  Thus
\begin{eqnarray}
&&\quad\mbox{Prob (go from $0$ to $n$ in time $t$) $=$} \nonumber \\
 \int^t_0   ds\,&&\mbox{Prob (go from $0$ to $0$ in time $t-s$)} \
 \cdot \ \gamma \ ds \cdot \nonumber    \\
&&\quad {}\cdot  \mbox{Prob (go from $\bar{1}$ to $n$ without hitting $0$
in time $s$)} \ \ .  
\label{eq:17new}
\end{eqnarray}
Using (\ref{eq:16new}) it follows that for some $\gamma s\leq \gamma t \leq
n^A$, 
\begin{equation}
\mbox{Prob (go from $\bar{1}$ to $n$ without hitting $0$ in time $s$)} \ \geq
\frac{1}{\gamma t n^B}
\geq \frac{1}{n^{A+B}} \ \ .
\label{eq:18new}
\end{equation}
However this last probability is the same for $T'_n$ as for $T_n$.  Turning to
the trees $T'_n$ we see that the node $0$ is connected to three other nodes, the
node at level $-1$ on the semi-infinite tree and the two nodes at level $1$.  In
time $\frac{1}{\gamma}$ there is an $n$-independent lower bound on the
probability of going from $0$ to $\bar{1}$.  Combining this fact with
(\ref{eq:18new}) we see that in a time $s+\frac{1}{\gamma}$ there is a
probability of going from $0$ to
$n$ on
$T'_n$  which is greater that $\frac{1}{n}$ to a power.  Thus if $\{ T_n \}$ is
classically penetrable so is $\{ T'_n
\}$.

We are now going to compare the classical and quantum evolution through the
family of trees $\{T'_n\}$.  From this point on we set $\gamma = 1$.We will
return to finite trees later in this section but for now the device of appending
a semi-infinite line to the trees of interest actually makes the analysis
simpler.  Again call the starting node (which is at level 0 of the tree $T'_n$)
$0$ and call the unique node at the
$n^{th}$ level $n$.  Then
\begin{equation}
p(t)=\langle n|e^{-\hat{H}t}|0\rangle
\label{eq:2.14}
\end{equation}
is the probability to go from $0$ to $n$ in time $t$ if you evolve with the classical rule. 
Similarly
\begin{equation}
A(t)=\langle n|e^{-i\hat{H}t}|0\rangle
\label{eq:2.15}
\end{equation}
is the quantum amplitude to be at  $n$ at time $t$ if at $t=0$ you are at $0$
and you evolve with the quantum Hamiltonian $\hat{H}$.  (Of course $\hat{H},p(t)$
and $A(t)$ are all sequences that depend on the sequence $\{T'_n\}$ but we will
not bother to place an $n$ label on these quantities.)

The Hamiltonian $\hat{H}$ is defined by (\ref{eq:3}) for each tree $T'_n$ but
now the number of nodes is infinite so the Hilbert space is infinite
dimensional.  
  Call the energy eigenvectors
$|E\rangle$ where
\begin{eqnarray}
\hat{H}|E\rangle &=& E|E\rangle  \nonumber \\
\noalign{\hbox{and}}
\langle E|E'\rangle &=&  \delta (E-E') 
\label{eq:2.16}
\end{eqnarray}
for the continuous part of the spectrum and
\begin{equation}
\langle E_r|E_s\rangle = \delta_{rs}\qquad \nonumber
\end{equation}
for the bound states. 
Now for any Hermitian operator $\hat{H}$, with matrix elements $H_{ab}$, any
eigenvalue $E$ of $\hat{H}$ must lie\cite{ref:4} in the union (over $a$) of the
intervals
\begin{equation}
|E-H_{aa}|\leq \sum_{b\neq a} |H_{ba}|
\label{eq:2.17}
\end{equation}
which, given the form (\ref{eq:3}), implies that the eigenvalues lie in the
interval~$[0,6]$.

Using the completeness of the $|E\rangle$'s we can write
(\ref{eq:2.14}) as
\begin{eqnarray}
p(t)&=&\int_0^6 \!  dE\, e^{-Et}\langle n|E\rangle\langle E|0\rangle
\label{eq:2.18}\\
\noalign{\hbox{and (\ref{eq:2.15}) as}}
A(t)&=&\int_0^6 \!  dE\, e^{-iEt}\langle n|E\rangle\langle E|0\rangle
\label{eq:2.19}
\end{eqnarray}
where the integral $dE$ is to be interpreted as a sum on the discrete part of the
spectrum.  From (\ref{eq:2.19}) we have
\begin{equation}
\frac{1}{2\pi} \int^\infty_{-\infty} \!\!  dt' \ e^{iwt'} \ A(t') = \int_0^6 dE \ \delta
(w-E) \langle n|E\rangle \langle E|0\rangle \ \ .
\label{eq:2.20}
\end{equation}
Multiply both sides by $e^{-wt}$ and integrate $dw$ from $0$ to $\infty$ to get,
for
$t > 0$,
\begin{equation}
\frac{1}{2\pi} \int^\infty_{-\infty} \!\!  dt' \frac{A(t')}{t-it'} = p(t)
\label{eq:2.21}
\end{equation}
which could have been obtained using the Cauchy integral formula.  Now in the
$|a\rangle$ basis $\hat{H}$ is real and symmetric and from (\ref{eq:2.15}) it
then follows that $A(t)=A^\ast(-t)$.  This allows us to write (\ref{eq:2.21}) as
\begin{equation}
p(t)=\frac{1}{\pi} \ \mbox{Re} \  \int^\infty_0 \!\!  \ dt' \frac{A(t')}{t-it'} \ \ .
\label{eq:2.22}
\end{equation}

We will now use (\ref{eq:2.22}) to show that if a family of trees $\{T'_n\}$ is
classically penetrable it is also quantum penetrable.   Pick some time $T$ and
let $\epsilon$ be the maximum of $|A(t')|$ for
$0\leq t'\leq T$.  Now 
\begin{eqnarray}
p(t) & = & \frac{1}{\pi} \ \mbox{Re} \ \left\{\int^T_0 \!\!  dt'\frac{A(t')}{t-it'} +
\int^\infty_T \!\!  dt' \frac{A(t')}{t-it'}\right\} \nonumber \\[2ex]
 & \leq &
\frac{\epsilon}{\pi} \int^T_0 \!\!    dt' \frac{1}{(t^2 + t^{'2})^{1/2}} +
\frac{1}{\pi}
 \left| \int^\infty_T \!\!  dt'\frac{A(t')}{t-it'}\right|  \label{eq:2.23} \\[2ex] 
& = &
\frac{\epsilon}{\pi} \
\mbox{ln} \ \left[\frac{(T^2+t^2)^{1/2}+T}{t}\right]+\frac{1}{\pi}  \left| \int^\infty_T
\!\!  dt' \frac{A(t')}{t-it'}\right| \ \ . \nonumber
\end{eqnarray}
The magnitude of the last integral in (\ref{eq:2.23}) is actually less than
$C/T^{1/4}$ for large $T$ where $C$ is an $n$-independent constant.  We will
show this shortly.  With this result we then have that
\begin{equation}
p(t) \leq \frac{\epsilon}{\pi} \ \mbox{ln} \
\left[\frac{(T^2+t^2)^{1/2}+T}{t}\right] +
\frac{C}{T^{1/4}}\ \ .
\label{eq:2.24}
\end{equation}
Now we are assuming that the family of trees is classically penetrable.  This
means that for some $t \leq n^A$  we have $p(t) > 1/n^B$ for some $A$ and $B$.  For
large
$n$, this penetration time $t$ is clearly $\geq 1$.  Since the $\ln$ term in
(\ref{eq:2.24}) is a decreasing function of $t$, we have
\begin{equation}
\frac{1}{n^B} \leq \frac{\epsilon}{\pi} \ln \left[(T^2 + 1)^{1/2} + T\right] 
+ \frac{C}{T^{1/4}} \ \ .
\label{eq:32new}
\end{equation}
Now let $T=n^D$ for $D>4B$.  We then have, for large enough $n$, 
\begin{equation}
\frac{1}{n^B} \leq \frac{\epsilon}{\pi} \ln (n^D)
\label{eq:33new}
\end{equation}
which means that the maximum of $|A(t)|$ for $t<n^D$ is bigger than a constant
times $1/n^{B+1}$  Thus we have the result that if a family of trees $\{T'_n\}$
is classically penetrable, it is also quantum penetrable.

Before   verifying  that the last integral in (\ref{eq:2.23})
is actually bounded as claimed, we need to establish some facts
about the eigenfunctions of $\hat{H}$.  Label the nodes on the semi-infinite line
of $T'_n$ by $j$ with
$j=0,-1,-2,\ldots$ so that $j=0$ is the starting node of $T_n$.  On the
semi-infinite line
\begin{equation}
\qquad\qquad \hat{H}|j\rangle = 2|j\rangle-|j+1\rangle-|j-1\rangle \qquad 
\mbox{for~} j \leq -1 \ \ .
\label{eq:2.27}
\end{equation}
The state $|\theta\rangle$ with $\langle j|\theta\rangle$ proportional to
$e^{ij\theta}$ is an eigenstate of (\ref{eq:2.27}) with energy
\begin{equation}
E(\theta)=(2-2\cos\theta)=4\sin^2\theta/2\ \ .
\label{eq:2.27a}
\end{equation}
  Now $e^{ij\theta}$ and
$e^{-ij\theta}$ correspond to the same energy but because of the finite
branching part of the tree ($T_n$, which is connected at $j=0$), only one linear
combination is an eigenfunction of the full $\hat{H}$,
\begin{equation}
\langle j|\theta\rangle=\frac{1}{(2\pi)^{1/2}}\left[e^{ij\theta}+R(\theta)
e^{-ij\theta}\right]
\label{eq:2.28}
\end{equation}
with $0\leq\theta\leq\pi$, and $R(\theta)$ is determined by the structure of
$T_n$.  Because
$\hat{H}$ in the node basis is real, (\ref{eq:2.28}) must be real up to an
overall $j$ independent phase. This implies that $R(\theta)$ is of the form
$e^{-2i\delta(\theta)}$, that is,
$|R(\theta)|=1$.  (The form (\ref{eq:2.28}) is an ``in"
state for scattering off of the tree $T_n$ at the end of the semi-infinite
line.  The fact that $|R(\theta)|=1$ is also a consequence of the unitarity of the
$S$ matrix.)  We can rewrite (\ref{eq:2.28}) as
\begin{equation}
\langle j|\theta\rangle = e^{-i\delta(\theta)} \frac{2}{(2\pi)^{1/2}} \cos (j\theta + \delta (\theta))
\label{eq:36new}
\end{equation}
and then absorb the phase in the definition of $|\theta\rangle$ to get
\begin{equation}
\langle j|\theta\rangle = \Bigl(\frac{2}{\pi}\Bigr)^{1/2} \cos (j\theta + \delta (\theta))\
\ .
\label{eq:37new}
\end{equation}
The states
$|\theta\rangle$ are a set of delta function normalized eigenstates, i.e.,
\begin{equation}
\langle \theta | \theta'\rangle = \delta (\theta-\theta')\ \ .
\label{eq:38new}
\end{equation}
We have introduced the states $|\theta\rangle$ because we could (fairly) easily
normalize them, that is, pick the coefficient in (\ref{eq:2.28}) so that
(\ref{eq:38new}) holds.  The continuous energy eigenstates $|E\rangle$ given by
(\ref{eq:2.16}) are proportional to the $|\theta \rangle$'s.  To maintain both
(\ref{eq:2.16}) and (\ref{eq:38new}) we have
\begin{equation}
|E\rangle = \Bigl(\frac{dE}{d\theta}\Bigr)^{-1/2} |\theta\rangle =
(4E-E^2)^{-1/4}|\theta\rangle
\label{eq:39new}
\end{equation}
where again $E=4\sin^2\theta/2$.  In the node basis on the
semi-infinite line we then have
\begin{equation}
\langle j|E\rangle = \Bigl(\frac{2}{\pi}\Bigr)^{1/2} \frac{1}{(4E-E^2)^{1/4}}
\cos (j\theta +
\delta(E)) \ ,\qquad 0\le E \le 4.
\label{eq:40new}
\end{equation}

We now describe the bound-state part of the spectrum. Return to the form of
$\hat{H}$, given by (\ref{eq:2.27}) on the semi-infinite line, and consider the
eigenfunctions
\begin{eqnarray}
\langle j | \alpha \rangle &=& (-1)^j e^{\alpha j}\qquad \alpha>0 \nonumber\\
%\noalign{\hbox{and}}
\langle j |\beta \rangle &=& e^{\beta j} \qquad \beta>0
\label{eq:40new2}
\end{eqnarray}
with energies $2+2\cosh\alpha$ and $2-2\cosh \beta$, respectively. Since we know
that the eigenvalues of the full $\hat H$ (including the tree) lie in $[0,6]$, we see that
there are no bound states of the form~$|\beta\rangle$ and any bound states of the
form $|\alpha\rangle$ have energies in the interval $[4,6]$. We have now fully
explored the solutions to $\hat H | E\rangle = E|E\rangle$ on the runway. Any
additional solutions, which may be nonzero in the tree, will vanish identically on the
runway and will play no role in any of our discussion.

Next we prove the required bound for the last integral in (\ref{eq:2.23}) The trusting
reader is invited to skip beyond (\ref{eq:47new}). 
 First note that
\begin{equation}
A(t') = \langle n|e^{-i\hat{H}t'}|0\rangle = \int^6_0 \!\!  dE \langle
n|E\rangle\langle E |0\rangle e^{-iEt'}
\label{eq:41new}
\end{equation}
where the integral in the range from~4 to~6 is actually a sum. 
The integral in (\ref{eq:2.23}) we wish to bound is (after dividing by~$i$)
\begin{eqnarray}
\int^\infty_T \!\!  dt'\frac{A(t')}{t'+it} &=&\int^\infty_T \!\!  dt'\int^6_0 \!\!  dE \langle
n|E\rangle \langle E|0\rangle e^{-iEt'}\frac{1}{t'+it} \nonumber \\[2ex]
&=& \int^\infty_T \!\!  dt'\int^6_0 \!\! dE \, \langle n|E\rangle
\langle E|0\rangle e^{-iEt'}\int^\infty_0 \!\!   d\mu\, e^{-\mu(t'+it)} \label{eq:42new}
\\[2ex] &=&\int^6_0 \!\!  dE \int^\infty_0 \!\!  d\mu\, \langle n|E\rangle \langle E |
0\rangle e^{-i\mu t} e^{-iET} e^{-\mu T} \frac{1}{\mu + iE} \ \ . \nonumber
\end{eqnarray}
Taking the absolute value we get
\begin{equation}
\Bigl|\int^\infty_T \!\!  dt'\frac{A(t')}{t'+it}\Bigr| \leq \int^\infty_0 \!\!  d\mu\,
e^{-\mu T} \int^6_0 dE \left| \langle n|E\rangle \right|
\frac{| \langle E|0\rangle|}{(\mu^2+E^2)^{1/2}}
\label{eq:43new}
\end{equation}
By the Cauchy-Schwarz inequality,
\begin{eqnarray}
\Bigl| \int^\infty_T \!\!  dt' \frac{A(t')}{t'+it} \Bigr| &\leq&
\int_0^\infty d\mu e^{-\mu T} \Bigl[ \int_0^6 dE' |\langle n|E'\rangle|^2 \Bigr]^{1/2}\,
\Bigl[ \int_0^6 dE\frac{| \langle E|0\rangle|^2}{\mu^2+E^2}\Bigr]^{1/2}\nonumber\\[2ex]
&=& \int_0^\infty d\mu e^{-\mu T} \Bigl[\int_0^4 dE\frac{| \langle
E|0\rangle|^2}{\mu^2+E^2} + \sum_r\frac{| \langle
E_r|0\rangle|^2}{\mu^2+E_r^2}
\Bigr]^{1/2}
\label{eq:45new}
\end{eqnarray}
using $\langle n|n\rangle = 1$. For $0\le E\le 4$, the matrix element $\langle
E|0\rangle$ is given by (\ref{eq:40new}) so we have $|\langle E|0\rangle|^2 \le
C_1/(4E-E^2)^{1/2}$ where~$C_i$ here and below are easily computable constants. Since
$\sum_r|\langle E_r|0\rangle|^2 \le 1$, and each $E_r\ge 4$, we have 
 \begin{equation}
 \Bigl|\int^\infty_T \!\!  dt'\frac{A(t')}{t'+it}\Bigr| \leq C_2 \int^\infty_0 \!\!  d\mu\,
 e^{-\mu T} \Bigl[\int^4_0 \!\!  
 \frac{dE}{(4E-E^2)^{1/2}(\mu^2 + E^2)} + \frac1{\mu^2+4^2}\Bigr]^{1/2} \ \ .
 \label{eq:44new2}
 \end{equation}
  The integral $dE$ in (\ref{eq:44new2}) is
\begin{equation}
\int^4_0 \!\!  dE \frac{1}{(4E-E^2)^{1/2}} \ \frac{1}{\mu^2+E^2}  = 
\int_0^\pi \!\! d\theta \frac1{\mu^2+(4\sin^2\theta/2)^2} 
< \int_0^\pi \!\! d\theta \frac1{\mu^2+(2/\pi)^4 \theta^4} 
 \leq \frac{C_3}{\mu^{3/2}} \ \ .
\label{eq:46new}
\end{equation}
Now the inequality (\ref{eq:44new2}) becomes
\begin{equation}
\Bigl|\int^\infty_T \!\!  dt \ \frac{A(t')}{t'+it} \Bigr| \leq 
C_2 \int^{\infty}_{0} \!\!  d\mu\, e^{-\mu T}\Bigl[
\frac{C_3}{\mu^{3/2}} + \frac1{\mu^2 + 4^2}  \Bigr]^{1/2}
\leq C_4 \int^{\infty}_{0} \!\!  d\mu\,
e^{-\mu T}
\frac{1}{\mu^{3/4}} \leq \frac{C_5}{T^{1/4}}
\label{eq:47new}
\end{equation}
which is the desired result.  This was the last step we needed in showing that if
$\{T'_n\}$ is classically penetrable then it is quantum penetrable.

Of course we are not ultimately interested in quantum evolving on the family of
infinite trees $\{T'_n\}$ because we only imagine building a quantum computer
with a finite number of building blocks.  However we now argue that if the family
$\{T'_n\}$ is quantum penetrable there is a closely related family of finite
trees
$\{T^f_n\}$ that is also quantum penetrable.  In fact
$T^f_n$ is obtained from
$T'_n$ by chopping off the semi-infinite line at some node that is far, but not
exponentially far as a function of $n$, from the node $0$.  Alternatively we can
view
$T^f_n$ as arising from $T_n$ by appending to $T_n$ at $0$ a finite
number of linearly connected nodes.

To understand when infinite and very long give rise to the same quantum
evolution consider an infinite line of nodes by itself with the Hamiltonian given by
(\ref{eq:2.27}).  In this case it is possible to explicitly evaluate the
amplitude to go from $j$ to $k$ in time~$t$:
\begin{equation}
\langle k|e^{-i\hat{H}t}|j\rangle = e^{-2i t} i^{(k-j)} J_{k-j}(2t)
\label{eq:48new}
\end{equation}
where $J_{k-j}$ is a Bessel function of integer order.  For fixed $t$ this
amplitude dies rapidly if $|k-j|$ is bigger than $2t$.  Imagine starting
at $j=0$ at $t=0$.  The quantum amplitude spreads out with speed $2$ (recall
that we have set $\gamma$=1).  Chopping off the infinite system at the nodes $\pm
L$ will not affect the evolution from $j=0$ as long as  
$L \gg 2 t$.

Return to the family of quantum penetrable trees $\{T'_n\}$.  These trees have
the property that starting at $0$, which is at the end of the semi-infinite
line, there is a substantial quantum amplitude for being at the node on the
$n^{th}$ level of the branching tree at a time $t \leq n^{\bar{A}}$ for a fixed
$\bar{A}$.  Lopping off the infinite tree at a node of order $(n^{\bar{A}})^2$
down from $0$ will not affect this result.  Thus the family of finite trees
$\{T^f_n\}$, which are obtained from the family of classically
penetrable  trees $\{T_n\}$ by adding a finite number of linearly connected
nodes, is quantum penetrable.

It is reasonable to ask why we bother with the family of infinite trees
$\{T'_n\}$ when we are only actually interested in finite trees.  Why didn't we
prove directly that the family of classically penetrable trees $\{T_n\}$ is also
quantum penetrable?  Of course the answer is we would have if we could have. 
The difficulty lies in the fact that for an arbitrary finite tree with an
exponential number of nodes there are an exponential number of energy
eigenvalues falling in a fixed interval and we were unable to establish the
requisite facts about the density of states needed for a proof.

Let us summarize the results of this section.  We started with a given 
family of trees $\{T_n\}$ that was assumed to be classically penetrable.  We then
constructed the closely related family of trees $\{T'_n\}$ that has a
semi-infinite line of nodes attached to the starting node of each $T_n$.  The
trees $\{T'_n\}$ are also classically penetrable.  Then, using the analytic
relationship between the classical probabilities and quantum amplitudes of
$\{T'_n\}$ we were able to prove that
$\{T'_n\}$ is quantum penetrable.  We also argued that cutting the semi-infinite
line at some node far from 0 cannot affect the
quantum penetrability as long as the distance to the cut is much greater than the
quantum penetration time.  Therefore the family
$\{T^f_n\}$ of trees that is made from $\{T_n\}$ by appending a long
(but finite) string of nodes to the starting node of each $T_n$ is quantum
penetrable if the original $\{T_n\}$ is classically penetrable.  Clearly $\{T_n\}$ and
$\{T^f_n\}$ are addressing precisely the same decision question. 
Therefore any problem that can be solved by classically random walking
through a decision tree can be solved by quantum evolving through a very closely
related tree.

\section{A Family of Trees that is Quantum,\protect\\
 but  Not Classically, Penetrable}

If we know enough about the structure of a family of trees we can decide if it
is classically penetrable and if it is quantum penetrable.  Here we will show
examples of families of trees that are quantum but not classically penetrable. 
We begin by discussing the calculations in the quantum case.  As in the last
section we consider a family of trees $\{T_n\}$ whose members have only one node
at the
$n^{th}$ level, called $n$.  This time we construct the family
$\{T^{\prime\prime}_n\}$ where each tree $T''_n$ has two semi-infinite lines of nodes,
one connected to the starting node of $T_n$, and the other semi-infinite
line of nodes attached to the node $n$ of $T_n$.  For calculational purposes we make
these two extra lines of nodes semi-infinite but ultimately we envisage making them of
length
$n$ to a power.

For convenience we redraw our trees so that the direct line of nodes from $0$ to
$n$ lies along the base.  In this way the tree depicted in \Figcite{Figii} with
two semi-infinite lines appended becomes that of \Figcite{Figiv}.  We use
``bush'' to denote a group of nodes coming out of a node on the base.
\Fig{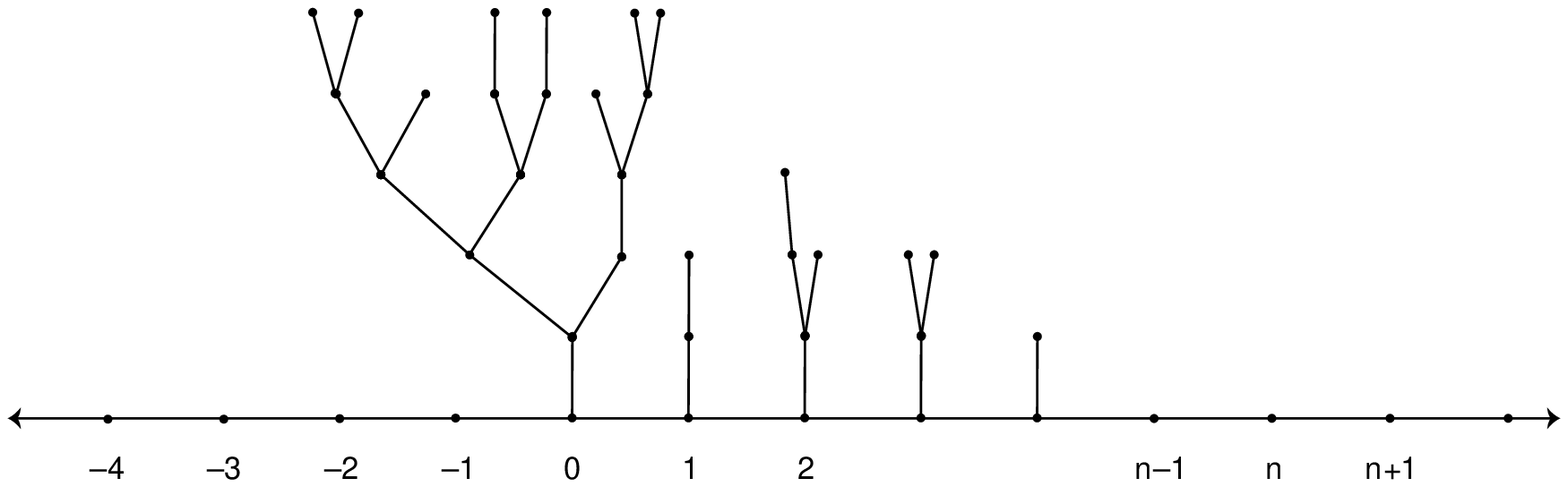}{900}{Figiv}{The tree $T''_n$ obtained from the tree $T_n$ of
Fig.\thinspace2 by appending two semi-infinite lines, one connected at the
starting node and one connected to the node~$n$. The tree is drawn with the direct line
of nodes from~$0$ to~$n$ along the base.}%
 Here we label the nodes on the base by $j$.  The nodes
$j=-1,-2,-3,\ldots$ are on the semi-infinite starting line.  The nodes
$j=n+1,n+2,\ldots$ are on the appended ending line.  The nodes $j=0,
\dotsc,n$ are all on the original tree $T_n$ and the nodes $0,\dotsc,n-2$ may
have bushes coming out them although the nodes $n-1$ and $n$ do not.  (If node
$n-1$ had a bush then $n$ would not be the unique level $n$ node.)  What we
imagine doing is building a quantum state localized near 0 on the starting line
and then calculating the quantum amplitude for penetrating the tree and being on
the ending line.  To this end we now set up the formalism for calculating the
energy dependent transmission coefficient $T(E)$ and then evaluate it in certain
specific cases of families of trees. 

For the tree depicted in \Figcite{Figiv} with an infinite base, for each energy $E$
with $0\leq E\leq 4$, there are two energy eigenstates.  (Here again we have
set $\gamma$ equal to 1).  On the semi-infinite lines they are, in the
node basis, of the form
$e^{ij\theta}$ and
$e^{-ij\theta}$ where again $E=4\sin^2\theta/2$ and $0 \leq \theta \leq \pi$. 
Superpositions of the
$e^{ij\theta}$ are used to make right moving packets whereas superpositions of
$e^{-ij\theta}$ make left movers.  Consider the state $|E,+ \mbox{in} \rangle$ 
that on the starting and ending lines is of the form
\begin{eqnarray}
\qquad\qquad\qquad \langle j|E,+ \mbox{in} \rangle & = & N(E) [e^{ij\theta} + R(E)
e^{-ij\theta}]
\qquad\, j=-1,-2, \dots
\nonumber \\
 \langle j|E,+ \mbox{in} \rangle & = & N(E) T(E)e^{ij\theta}
\phantom{[e^{ij\theta} +  1a}
\qquad j=n-1,n,n+1,\ldots \label{eq:49} \\
\noalign{\hbox{with}} 
N(E) &=& \frac{1}{(2\pi)^{1/2}} \, \frac{1}{(4E-E^2)^{1/4}} \ \ . \nonumber 
\end{eqnarray}
At this point we say nothing about $\langle a|E, + \mbox{in} \rangle$ if $a$ is
a node on $T_n$.  Superpositions of $|E,+ \mbox{in} \rangle$ make states that at
early times represent right moving packets on the starting line headed towards
the tree structure
$T_n$.  At late times the packet splits into a reflected piece, proportional
to $R$, left moving on the starting line, and a transmitted piece, proportional
to
$T$, which is right moving on the ending line. 
Similarly we can define  $|E, - \mbox{in} \rangle$, which represents a
state left moving on the ending line at early times that at late times is
split into a right mover on the ending line and a transmitted part left moving
on the starting line. For $|E, - \mbox{in} \rangle$ we have 
\begin{eqnarray}
\qquad\qquad\qquad\langle j|E,- \mbox{in} \rangle & = & N(E) [e^{-ij\theta} +
\bar{R}(E) e^{ij\theta}]
\qquad  j=n-1,n, n+1, \ldots
\nonumber \\
\langle j|E,- \mbox{in} \rangle & = & N(E) \bar{T}(E)e^{-ij\theta}
\phantom{(E) e^{ij\theta}]}
\qquad j=-1,-2,\ldots \  \ . 
\label{eq:50}
\end{eqnarray}
The states $|E, + \mbox{in} \rangle$ and $ |E, - \mbox{in}
\rangle$ are a complete set of scattering states useful for discussing tree
penetration.  Equivalently there is the set $|E, + \mbox{out} \rangle$
and $|E, - \mbox{out} \rangle$ that at late times represents
respectively a right mover on the ending line and a left mover on the starting
line.  From (\ref{eq:49}) and (\ref{eq:50}) we get
\begin{eqnarray}
|E, + \mbox{in} \rangle &=& R(E) | E, - \mbox{out} \rangle + T(E) |E,
+ \mbox{out} \rangle \nonumber \\
|E, - \mbox{in} \rangle &=& \bar{R}(E) | E, + \mbox{out} \rangle +
\bar{T}(E) |E, - \mbox{out} \rangle \ \ .
\label{eq:51}
\end{eqnarray}
This transformation from the out states to the in states is called the
$S$-matrix,
\begin{equation}
S= \left( \begin{array}{cc} 
R & T \\
\bar{T} & \bar{R}
\end{array} \right)
\label{eq:52}
\end{equation}
which is necessarily unitary so we have
\begin{eqnarray}
| R(E)|^2 &+& |T(E)|^2 = 1 \nonumber \\
| \bar{R}(E)|^2 &+& |\bar{T}(E)|^2 = 1 \label{eq:53}  \\
R^*(E)T(E) &+& \bar{T}^*(E)\bar{R}(E) = 0  \nonumber  \ \ . 
\end{eqnarray}

The standard interpretation of $T(E)$ is as follows.  Suppose we build a state
$|\psi \rangle$ completely on the starting line, that is, $\langle a |
\psi
\rangle$ is nonzero only for nodes $a$ on the starting line.  Furthermore suppose that $|\psi \rangle$ expanded as a
superposition of energy eigenstates is made only of states whose energy is
close to some $E_0$.  If we quantum mechanically evolve $|\psi \rangle$ with
the unitary operator $e^{-i\hat{H}t}$, then at late times the probability of
being on the ending line is $|T(E_0)|^2$.  Thus $|T(E)|^2$ has a direct
interpretation as the $E$ dependent transmission probability through the  tree.

Of course any state $|\psi \rangle$ that is highly localized in energy is
necessarily highly delocalized in the node basis.  (This can be viewed as a
consequence of the uncertainty principle.)   We don't want our constructions to
rely on building states that are very spread out on the starting line since we
eventually do wish to chop it off not too far from the node~$0$.   Suppose we
start at a specific node, $j$ on the starting line, and we want the amplitude
for being at node $k$ on the ending line at time~$t$.  This is given by
\begin{eqnarray}
A_{kj}(t) &=& \langle k|e^{-i\hat{H}t}|j\rangle \nonumber \\
 &=& \int^4_0 \!\!  dE \left\{ \langle k|E, + \mbox{in} \rangle \langle E, +
\mbox{in}|j \rangle + \langle k|E, - \mbox{in} \rangle \langle E, -
\mbox{in}|j \rangle \right\}e^{-iEt} + \sum_r \langle k|E_r\rangle \langle E_r|j\rangle
e^{-iE_r t}   \nonumber \\ 
&=& \int^4_0 \!\!  dE N^2(E)\left\{T(E)e^{ik\theta} (e^{-ij\theta} +R^*(E)e^{ij\theta}) +
(e^{-ik\theta} +\bar{R}(E)e^{ik\theta}) \bar{T}^*(E)e^{ij\theta} \right\}
e^{-iEt}\nonumber\\
&&\qquad{} + 
\sum_r \langle k|E_r\rangle \langle E_r|j\rangle e^{-iE_r t} \label{eq:54}
\end{eqnarray}
where we have used the explicit forms for $|E, \pm \mbox{in} \rangle$
on the starting and ending lines and also included possible bound states.
Now using the last equation in (\ref{eq:53}), with the further fact that $\hat H$ being
real in the node basis implies $T(E)=\bar T(E)$, we get
\begin{equation}
A_{kj}(t) = \int^4_0 \!\! \!  dE\, N^2(E) \Bigl\{T(E) e^{i(k-j)\theta} +
T^*(E)e^{-i(k-j)\theta}\Bigr\} e^{-iEt} + \! \sum_r \langle k|E_r\rangle \langle
E_r|j\rangle e^{-iE_r t}.
\label{eq:55}
\end{equation}
In order to obtain amplitudes~$A_{kj}$ that are large enough to ensure penetrability,
we will look for trees for which $T(E)$ is large and non-oscillatory in some
interval of $E$'s. This guarantees that the right-hand side of (\ref{eq:55}) is large
enough at some relevant time.

We now turn to calculating $T(E)$, which clearly depends on the structure of the
tree to which we have added the semi-infinite starting and ending lines of
nodes.  For each of the nodes $m=0,1,\dotsc, n-2$ along the base of the tree --
see \Figcite{Figiv} -- that has a bush sprouting up from it, let us define
\begin{equation}
y_m(E) = \frac{\langle {\rm node~above}~m|E, + {\rm in} \rangle}{\langle m|E,
+ {\rm in} \rangle}
\label{eq:56}
\end{equation}
where $|$node above $m\rangle$ is the state corresponding to the node one level
up from the base above the node $m$.  Now for fixed $E$, $y_m(E)$ is determined
solely by the bush coming out of the node $m$; it does not depend on the
other bushes.  To see this suppose that the bush coming out of node $m$ has
$N$ nodes above the base node $m$.  Label these nodes by $a=1, \dotsc, N$.  Now
$\hat{H}|a \rangle$ gives a superposition of $|a \rangle$ and the states
connected to $a$.  Thus
\begin{equation}
\langle a |\hat{H}|E, + {\rm in} \rangle = E \langle a | E, + {\rm in}\rangle
\label{eq:57}
\end{equation}
is $N$ equations for the $(N+1)$ quantities $\langle a | E, + {\rm in}
\rangle$ and $\langle m | E, + {\rm in} \rangle$.  Divide through by
$\langle m | E, + {\rm in} \rangle$ and we get $N$ equations for the $N$
ratios $\langle a | E, + {\rm in} \rangle/\langle m | E, + {\rm
in} \rangle$ so we see that (\ref{eq:56}) is determined by the bush alone. 
Furthermore the equations that were used to determine $y_m(E)$ are all real so
$y_m(E)$ is also real. For any given bush $y_m(E)$ can be calculated recursively by
looking at sub-bushes and it is not actually necessary to solve the~$N$ equations
(\ref{eq:57}).

Let $m$ be a node on the base with a bush coming off.  Now, from
(\ref{eq:3}),
\begin{eqnarray}
\langle  m|\hat{H}|E, + {\rm in} \rangle &=& 3 \langle m|E, + {\rm in}
\rangle \nonumber \\
 && - \langle  m+1|E, + {\rm in} \rangle - \langle  m-1|E, + {\rm in}
\rangle - \langle {\rm node~above~}m|E, + {\rm in} \rangle \nonumber \\ 
&=& E \langle  m|E, + {\rm in} \rangle
\label{eq:58}
\end{eqnarray}
which implies that 
\begin{equation}
\langle  m+1|E, + {\rm in} \rangle = (3-E-y_m(E)) \langle m|E, + {\rm in} \rangle
- \langle  m-1|E, + {\rm in} \rangle
\label{eq:59}
\end{equation}
where we have used (\ref{eq:56}).  If $m$ has no bush coming out of it, a
parallel argument gives
\begin{equation}
\langle  m+1|E, + {\rm in} \rangle = (2-E) \langle m|E, + {\rm in} \rangle
- \langle  m-1|E, + {\rm in} \rangle \ \ .
\label{eq:60}
\end{equation}
We can use (\ref{eq:59}) for nodes with bushes as well as without if we
define $y_m(E) =1$ for nodes on the base with no bushes above. 
Equation~(\ref{eq:59}) can be written as a matrix equation
\begin{equation}
\left[
\begin{array}{c}
\langle  m+1|E, + {\rm in} \rangle \\
\langle m|E, + {\rm in} \rangle
\end{array}
\right] =
\left[
\begin{array}{cc}
(3-E-y_m(E)) &\quad -1 \,\, \\
1 & 0
\end{array}
\right] 
\left[
\begin{array}{c}
\langle m|E, + {\rm in} \rangle \\
\langle  m-1|E, + {\rm in} \rangle
\end{array}
\right] \ \ .
\label{eq:61}
\end{equation}
We then have
\begin{equation}
\left[
\begin{array}{c}
\langle  n|E, + {\rm in} \rangle \\
\langle n-1|E, + {\rm in} \rangle
\end{array}
\right] = M
\left[
\begin{array}{c}
\langle  0|E, + {\rm in} \rangle \\
\langle -1|E, + {\rm in} \rangle
\end{array}
\right]  
 \ \ .
\label{eq:62}
\end{equation}
where
\begin{equation}
M=M_{n-1} \, M_{n-2} \cdots M_0
\label{eq:63}
\end{equation}
and
\begin{equation}
M_m = \left[
\begin{array}{cc}
(3-E-y_m(E)) &\quad -1 \,\, \\
1 & 0
\end{array}
\right]  
 \ \ .
\label{eq:62A}
\end{equation}
Substituting the explicit form for $|E, + {\rm in} \rangle$ from (\ref{eq:49})
we get
\begin{equation}
\left[
\begin{array}{c}
T(E) e^{in\theta} \\
T(E) e^{i(n-1)\theta}\end{array}
\right] = M
\left[
\begin{array}{c}
1 + R(E) \\
e^{-i\theta} + R(E)e^{i\theta} 
\end{array}
\right]  
 \ \ .
\label{eq:63A}
\end{equation}
If we know the matrix $M$, $T(E)$ is determined by these last two equations for
$T(E)$ and $R(E)$.  From (\ref{eq:62A}) we see that $M$ is the
product of matrices of determinant 1 so $\det (M)=1$. We can write
\begin{equation}
M =
\left[
\begin{array}{cc}
a & b \\
c & d
\end{array}
\right]  
\label{eq:64}
\end{equation}
with $ad-bc=1$ and $a,b,c,d$ all real.  Solving for $T(E)$ we get
\begin{equation}
T(E) = e^{-in\theta} \frac{2i \sin \theta}{c-b + (d-a) \cos \theta + i (d+a)
\sin \theta} \ \ .
\label{eq:3.19}
\end{equation}

It is interesting to note that if for some $E$ we have $y_m(E) = 1$ for all~$m$, then
$T(E)=1$. To see this we construct $M=M(E)$ in this special case. From~(\ref{eq:63})
and~(\ref{eq:62A}) we have
\begin{eqnarray}
M(E) &=& \biggl[\begin{array}{cc}
             2-E &\quad -1\\
               1  & 0 \end{array}\biggr]^n\nonumber\\
 &=& \frac1{\sin(\theta)} \biggl[\begin{array}{cc}
             \sin\bigl((n+1)\theta\bigr) & -\sin(n\theta)\\
             \sin(n\theta)  & -\sin\bigl((n-1)\theta\bigr) \end{array}\biggr]\ \ .
\end{eqnarray}
Plugging into (\ref{eq:64}) and (\ref{eq:3.19}) we get $T(E)=1$. To understand why this
comes about recall that a node with no bush is the same as a node with a bush for
which 
$y_m (E) = 1$ as far as the calculation of $T(E)$ is concerned. Therefore if all
bushes have $y_m (E) = 1$ at some $E$ we have unimpeded transmission at that~$E$.

To recap, given a decision tree $T_n$ with one node at level $n$, construct a
new tree with semi-infinite lines attached to the starting node 0 and to the node at
level $n$.  Redraw the tree as in \Figcite{Figiv} with the direct line
from 0 to $n$ along the base.  Suppose we can calculate the $n-1$ functions
$y_0(E), y_1(E), \dotsc, y_{n-2}(E)$.  Substitute into (\ref{eq:62A}) and
(\ref{eq:63}) to get the matrix $M$ as a function of $E$.  The transmission
coefficient $T(E)$ is then given by (\ref{eq:3.19}) where $E=4\sin^2 \theta/2$.

In order for a family of trees to be quantum penetrable, the function $|T(E)|$
must be not too small over a not too small range of $E$ as can be seen from
(\ref{eq:55}).  Furthermore even if $|T(E)|$ is not small, $T(E)$ must not
oscillate rapidly about zero or else the integral in (\ref{eq:55}) may be small due
to cancellations.  It is interesting to note that for any tree $T(E) \to 1$ as $E
\to 0$.  To see this note that the zero-energy eigenvector of
$\hat{H},|E=0,+\mbox{in}\rangle$, is constant in the node basis, that is, $\langle
a|E=0,+\mbox{in}\rangle$ is independent of $a$.  Thus $y_m(0)$ defined by
(\ref{eq:56}) is 1 for all nodes on the base and by the argument of the 
paragraph before last we have 
$T(0)=1$.  For trees that are not quantum penetrable we will see that although
$T(0)=1$, $T(E)$ falls to near zero at an exponentially small value of~$E$.

Consider a decision tree that is perfectly bifurcating until level $n-1$ and
then only one of the $2^{n-1}$ nodes at level $n-1$ continues on to level $n$. 
The associated tree $T_n$ is shown in~\Figcite{Figv}.  This decision tree could arise
from the following question.  You are given a list of $N=2^{n-1}$ items with the
knowledge that a single unspecified item may or may not be marked.  The
question is, ``Is there a marked item?"  (This is essentially the problem
for which Grover~\cite{ref:5} found a quantum algorithm requiring order $\sqrt N$
steps.)  Any classical algorithm for solving this problem requires of order $N$ steps.  In
particular the Markov process for moving through the decision tree gives a probability
of being at the unique node at level $n$ that is at most of order $1/N$, so this family
of trees is classically impenetrable.

\Fig{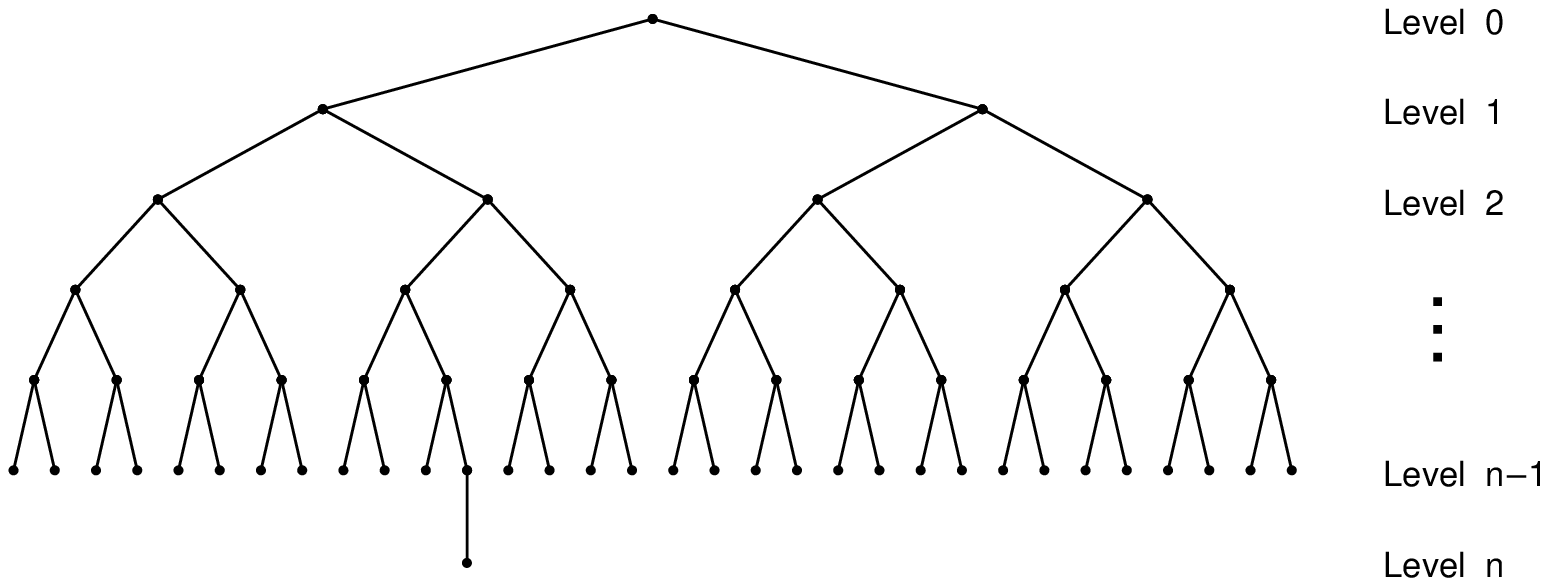}{900}{Figv}{The tree, $T_n$, which is perfectly bifurcating for the
first $n-1$ levels and then has only one node at level~$n$.}%

\Fig{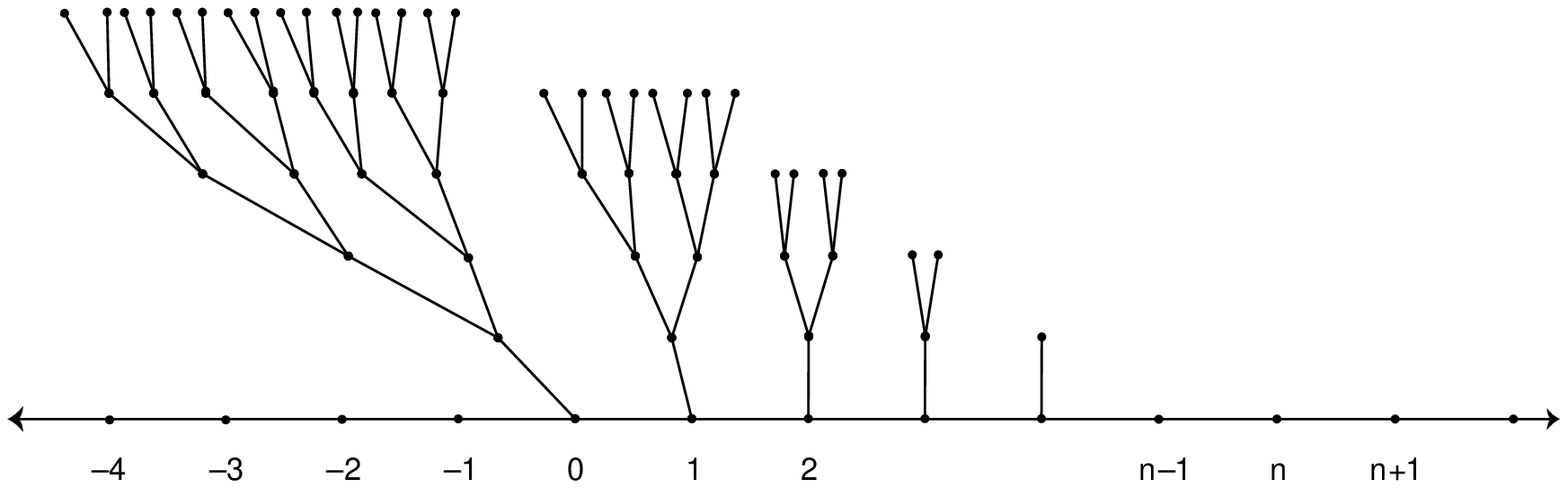}{900}{Figvi}{The tree $T''_n$ constructed from $T_n$ of
Fig.\thinspace5 by appending two semi-infinite lines of nodes and drawing the direct
line of nodes from~$0$ to~$n$ along the base.}

We now turn to quantum evolution through the same set of trees.  Draw the tree
in \Figcite{Figv} with the direct line from $0$ to $n$ along the base and add
semi-infinite starting and ending lines; see \Figcite{Figvi}.  We see that each bush
coming out of the base at node $m$ is a perfectly bifurcating bush of length $n-1-m$
for
$m=0$ to $n-1$.  The ratio $y_m(E)$ can be calculated for each
of these bushes.  Consider one such bush of length $k=n-1-m$ as depicted in
\Figcite{Figvii}.
\Fig{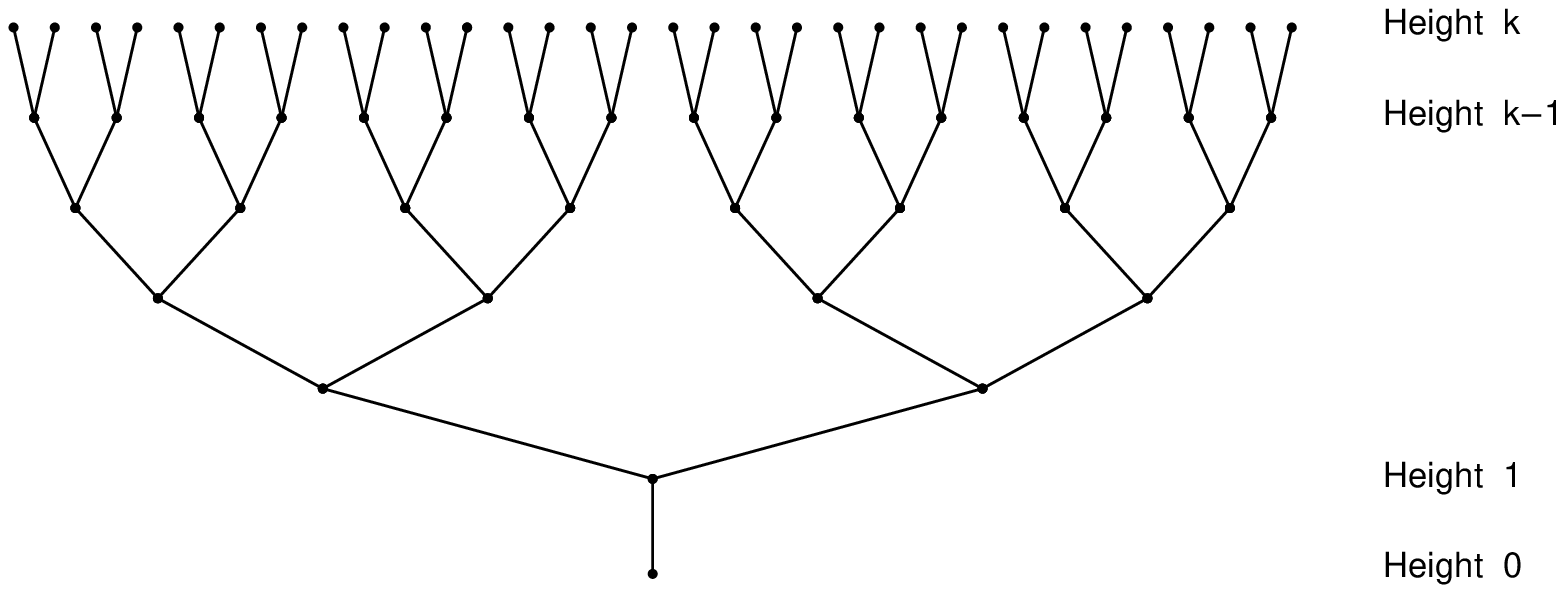}{900}{Figvii}{A perfectly bifurcating bush of height~$k$ coming out
of the base of the tree in Fig.\thinspace6 at node $m=n-1-k$.}%
At height $\ell$, with $1\leq \ell \leq k$, there are $2^{\ell -1}$ nodes.  At each
height we define the normalized state
\begin{equation}
|\ell ; pb\rangle = \frac{1}{(2^{\ell -1})^{1/2}}  
\sum_{a~{\rm at\  height}~\ell} \ |a\rangle
\label{eq:3.21}
\end{equation}
with $|0;pb\rangle$ being the state at the node on the bottom of the bush,
that is, $|0;pb\rangle=|m\rangle$.  With these labels, for these bushes,
$y_m(E)$ defined by (\ref{eq:56}) is
\begin{equation}
y_m(E) = \frac{\langle 1;pb|E,+ \mbox{ in}\rangle}{\langle 0; pb |E,+
\mbox{ in}\rangle} \ \ .
\label{eq:3.22}
\end{equation}
Note that $\hat{H}$ to any power acting on $|0;pb\rangle$ gives a linear
superposition of states that only contains the states $|\ell ; pb\rangle$ on
the bush.  Further note that
\begin{equation}
\qquad\langle \ell ; pb|\hat{H}|\ell ';pb\rangle = 3 \delta_{\ell\ell '} - \sqrt{2} 
[\delta_{\ell,\ell ' +1} + \delta_{\ell,\ell '-1}] \qquad  \mbox{for $1 \leq \ell$, 
$\ell ' \leq k-1$}
\label{eq:3.23}
\end{equation}
so the bush in \Figcite{Figvii} can be replaced by the effective linear bush given
in \Figcite{Figviii} where the number next to the node on the right gives the diagonal
element of the Hamiltonian and the number by the connecting edge on the
left gives the off-diagonal element.  Up to an overall constant that drops out of
\ref{eq:3.22}, for 
$\ell=1$ to
$k$ we have
$$\langle \ell ; pb|E,+   \mbox{ in}\rangle = \cos(\ell\theta'+\alpha)$$
and
\begin{equation}
\langle 0;pb|E, +{\rm in}\rangle = \sqrt{2} \cos \alpha
\label{eq:3.24}
\end{equation}
with
$$
 E=3-2\sqrt{2} \cos\theta'\ \ .
$$
\Fig{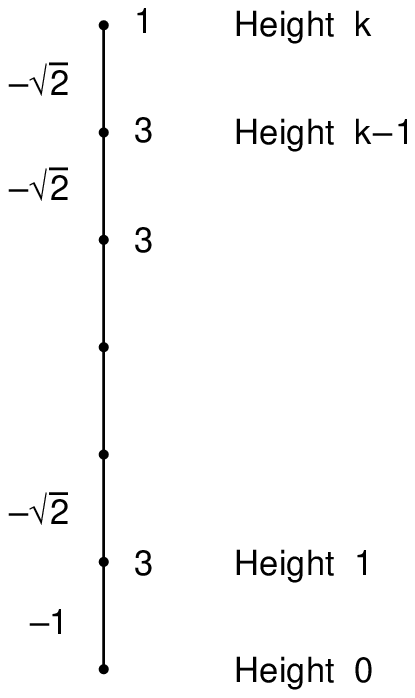}{900}{Figviii}{The effective bush of height~$k$ associated with the
bush of Fig.\thinspace7. The number to the left of each edge gives the matrix element
of~$\hat H$ between the two states connected by the edge. The number next to the
node gives the diagonal element of~$\hat H$ for that state.}%
By applying $\hat{H}$ to the $\ell=k$ node we can determine $\alpha$,
\begin{equation}
\tan (k\theta'+\alpha)=\frac{\cos(\theta')-\sqrt{2}}{\sin\theta'} \ \ .
\label{eq:3.24A}
\end{equation}
Going back to (\ref{eq:3.22}) we then have
\begin{equation}
y_m(E) = \frac{1}{\sqrt{2}}
\left\{\frac{\sqrt{2}\sin((k-1)\theta')-\sin(k\theta')}
{\sqrt{2}\sin(k\theta')-\sin((k+1)\theta')} \right\}
\label{eq:3.24B}
\end{equation}
where again $k=n-1-m$. Of course the calculation of $y_m(E)$ in this example was
greatly facilitated by the regularity of the bush.

With $y_m(E)$ determined for each bush we can evaluate $T(E)$ by substituting
into (\ref{eq:62A}), (\ref{eq:63}) and then
(\ref{eq:3.19}).  In \Figcite{Figix} we show $|T(E)|$ for $n=26$.  At the $n-1$ level
there are $2^{25}=10^{7.5}$ nodes.  Although $T(0)=1$, $T(E)$ has fallen
substantially by $E=10^{-10}$.  Most of the area under the curve comes from $E$
of order 1.  We can evaluate $T(E)$ explicitly at $E=3$.  Note from
(\ref{eq:3.24}) that $\theta ' = \pi/2$ at $E=3$.  In this case $y_m(3)$ is 1
if $k=n-1-m$ is even and $y_m(3)$ is $-1/2$ if $k$ is odd.  Thus $M(3)$ can be
written as (for~$n$ even)
\Fig{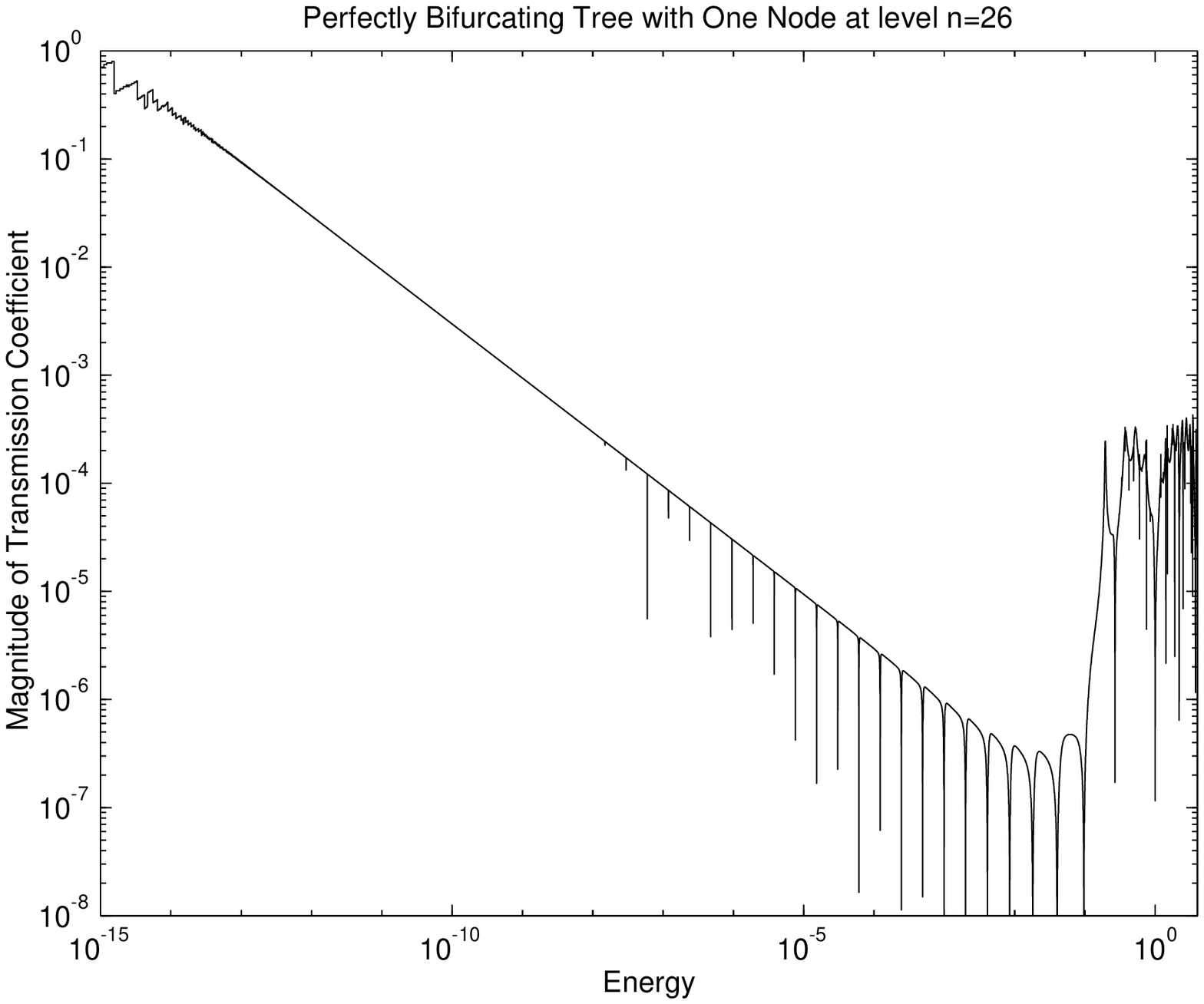}{900}{Figix}{The magnitude of $T$ versus $E$ for $E$ between~$0$
and~$4$ for the perfectly bifurcating tree with one node at the $n^{\rm th}$ level.}%
\begin{eqnarray}
M(3) & = & \left\{\left[\begin{array}{cc}1/2 & -1 \\ 1 & 0
\end{array}\right] \ \left[\begin{array}{cc}-1 & -1 \\ 1 & 0
\end{array}\right] \right\}^{n/2} \nonumber \\
& = & (-1)^{n/2} \left[\begin{array}{cc} 3/2 & 1/2 \\ 1 & 1 \end{array}\right]
^{n/2}
\\ & = & (-1)^{n/2} \left[\begin{array}{cc}1 & -1/3 \\ 1 & 2/3
\end{array}\right]\left[\begin{array}{cc} 2^{n/2} & 0 \label{eq:3.27} \\
 0 &
2^{-n/2}
\end{array}\right]\left[\begin{array}{cc} 2/3 & 1/3 \\ -1 & 1 \end{array}\right]
\nonumber
\end{eqnarray}
from which we conclude that $T(3)\sim 2^{-n/2}$.  The transmission amplitude is
of order $2^{-n/2}$ so the transmission probability goes like $2^{-n}$.  Here
the quantum algorithm is doing no better than the classical algorithm.

The alert reader may wonder whether any use can be made of the bound states
which may exist for $4\le E\le 6$. The answer is no, at least in this case. To check this,
we  changed the Hamiltonian on the semi-infinite lines to have values~3 on the
diagonal and $-$3/2 between neighbors. Now the continuum states $|E,\pm{\rm
in}\rangle$ are defined for $0\le E\le 6$ and are complete. We recalculated $T(E)$ and
looked for intervals of $E$'s where $T(E)$ is large and nonoscillatory. Again, there are
no values of $T(E)$ which permit transmission with probability greater
than~$\sim2^{-n}$.

Now we make a seemingly small modification of the tree.  We take all of the odd-height
bushes coming out of the base line of \Figcite{Figvi} and trim  back one layer so
all bushes are of even height.  The magnitude of the transmission coefficient is shown
in \Figcite{Figx} where we see that for a substantial range of
$E$ near 3, $|T(E)|$ is very close to~1. In fact for
all of these  teeth, $y_m(3)=1$, which by the argument given above implies that
$T(3)=1$.  We can also see that $T(E)$ does not oscillate rapidly in this region by
plotting the real part of $T(E)$, which is shown in \Figcite{Figxi}, confirming a more
tedious analytic evaluation.  Therefore the family of trees is quantum penetrable. 

\Fig{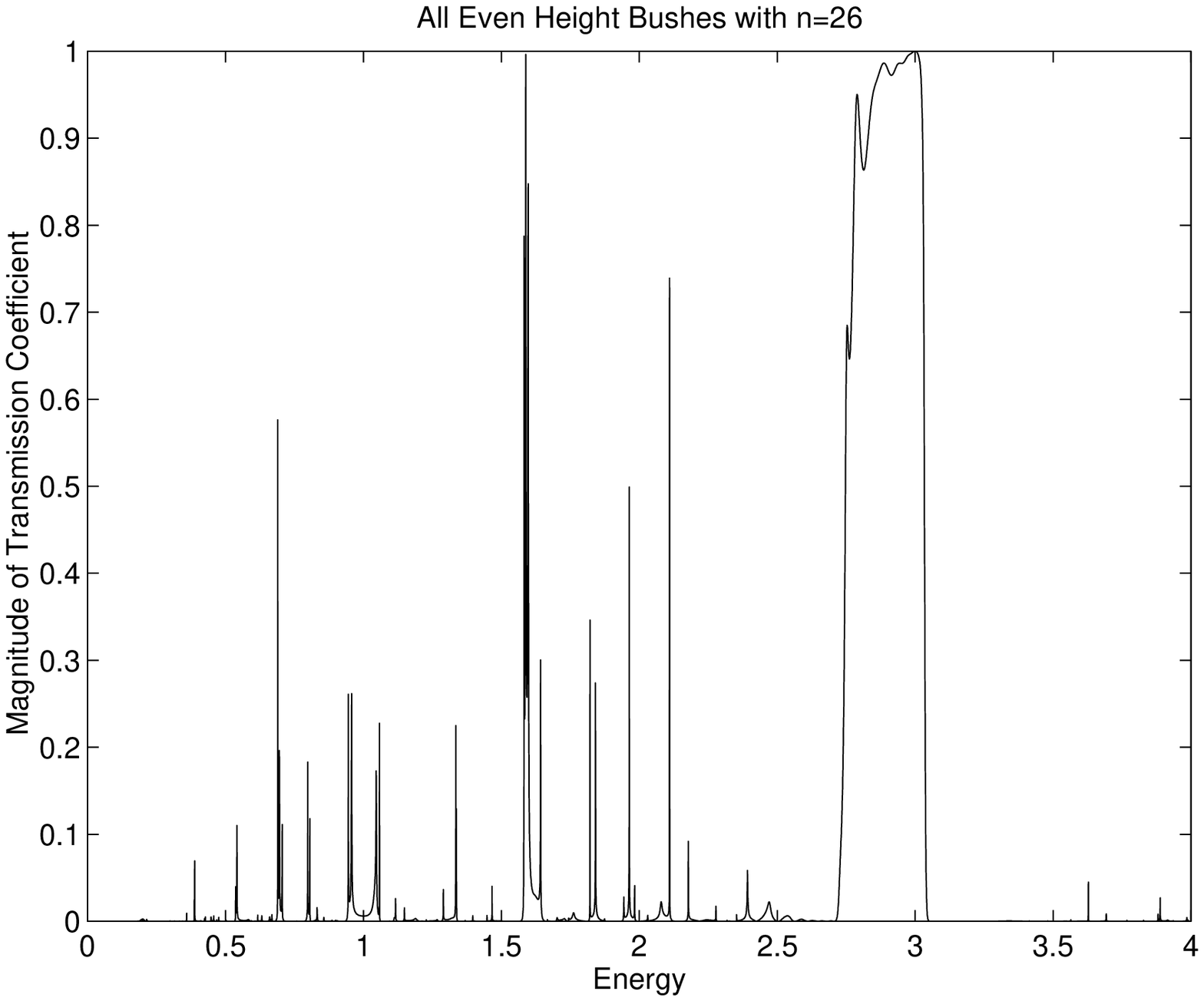}{900}{Figx}{The magnitude of $T$ versus $E$ for the same tree
used in Fig.\thinspace9 after removing one layer of nodes from each odd-length
bush.}%

It is easy to see that these trees with even-height bushes are not classically
penetrable. Before trimming back the odd-height bushes we had the $n$-level tree
shown in \Figcite{Figv}, $T_n$, which is associated with the  tree $T''_n$ shown in
\Figcite{Figvi}. These trees are not classically penetrable. Now, if we trim the
odd-height bushes back one layer, the trimmed tree still contains all of the tree
$T''_{n-1}$, which has even- and odd-height bushes. Since $T''_{n-1}$ is not classically
penetrable, the even-height bush family is also not classically penetrable, since,
classically, any time you add nodes to bushes you necessarily decrease the chances of
getting to the node~$n$. 

We have given a single example of a family of trees that is not classically penetrable
but {\it is\/} quantum penetrable. Clearly there are many variants of this example
using even-length, perfectly bifurcating bushes in all sorts of combinations; we will
not pursue these other examples here. However, we are faced  with the question of
what problem this family of trees corresponds to. 

We can think of decision trees as associated with functions that impose constraints. At
each level~$i$ there is a function $f_i$ that depends on the first $i$~bits.  If
$f_i(x_1\cdots x_i)=1$ then the $i^{\rm th}$-level node $x_1\cdots x_i$ is connected
to the $(i-1)^{\rm th}$-level node $x_1\cdots x_{i-1}$. (The $0^{\rm th}$-level node
needs no bits to describe it.) If $f_i(x_1\cdots x_i)=0$ then $x_1\cdots x_i$ is
absent from the tree. In terms of the functions $f_i$, the decision question is, ``Is
there  an $x_1\cdots x_n$ such that $f_i(x_1\cdots x_i)=1$ for all $i=1$ to~$n$?"

For the tree depicted in \Figcite{Figv}, the functions $f_1, \dotsc, f_{n-1}$ are all
identically~$1$. This gives the perfectly bifurcating structure. Then there is a function
$f_n(x_1\cdots x_n)$ that is guaranteed to be~$0$ for all but one of the $2^n$ values of
$x_1\cdots x_n$. At one special, but unknown, value $f_n$ is either $0$ or~$1$. (We
draw the decision tree assuming there is a value for which $f_n$ equals~$1$.
Otherwise the transmission coefficient is~$0$ and there is nothing to calculate.)
Without further information about $f_n$, any classical algorithm will need to search
$2^n$ values of $x_1\cdots x_n$ to see if there is a value at which $f_n$ equals~$1$.

Let us turn to the functions that determine the quantum penetrable tree just
discussed. At the $n^{\rm th}$ level there is the function $f_n(x_1\cdots x_n)$ which
may take the value~$1$ on one input, say $w_1\cdots w_n$. To arrange for the bushes
to all have even height, the tree must be trimmed at level~$n-1$. For $n$ even, the
function $f_{n-1}(x_1\cdots x_{n-1})$ is~$0$ if $x_1\neq w_1$ or if $x_1=w_1$,
$x_2=w_2$, and $x_3\neq w_3$ or if $x_1=w_1$, $x_2=w_2$, $x_3=w_3$, $x_4=w_4$,
and $x_5\neq w_5$, etc. If we are allowed to call the function $f_{n-1}(x_1\cdots
x_{n-1})$, which we know has this much structure, we can determine (thanks to
M.~Sipser) 
$w_1\cdots w_{n-1}$ with far fewer than order~$2^n$ function calls. First try various
 inputs until you find an example $x_1\cdots x_{n-1}$ such that $f_{n-1}$ is~$1$ on this
input. Then you know that $w_1=x_1$. Trying inputs of the form $w_1x_2\cdots
x_{n-1}$ will allow you to find $w_2$, etc. Once $w_1\cdots w_{n-1}$ is determined,
two function evaluations of $f_n(w_1\cdots w_{n-1} x_n)$ with $x_n=0,1$ will answer
the decision question. Of course what is occurring here is that the extreme regularity of
the tree, which guarantees its quantum penetrability, is also structuring the decision
problem so that it can be answered much more efficiently than by a classical random
walk, which is incapable of seeing larger structures. 

\Fig{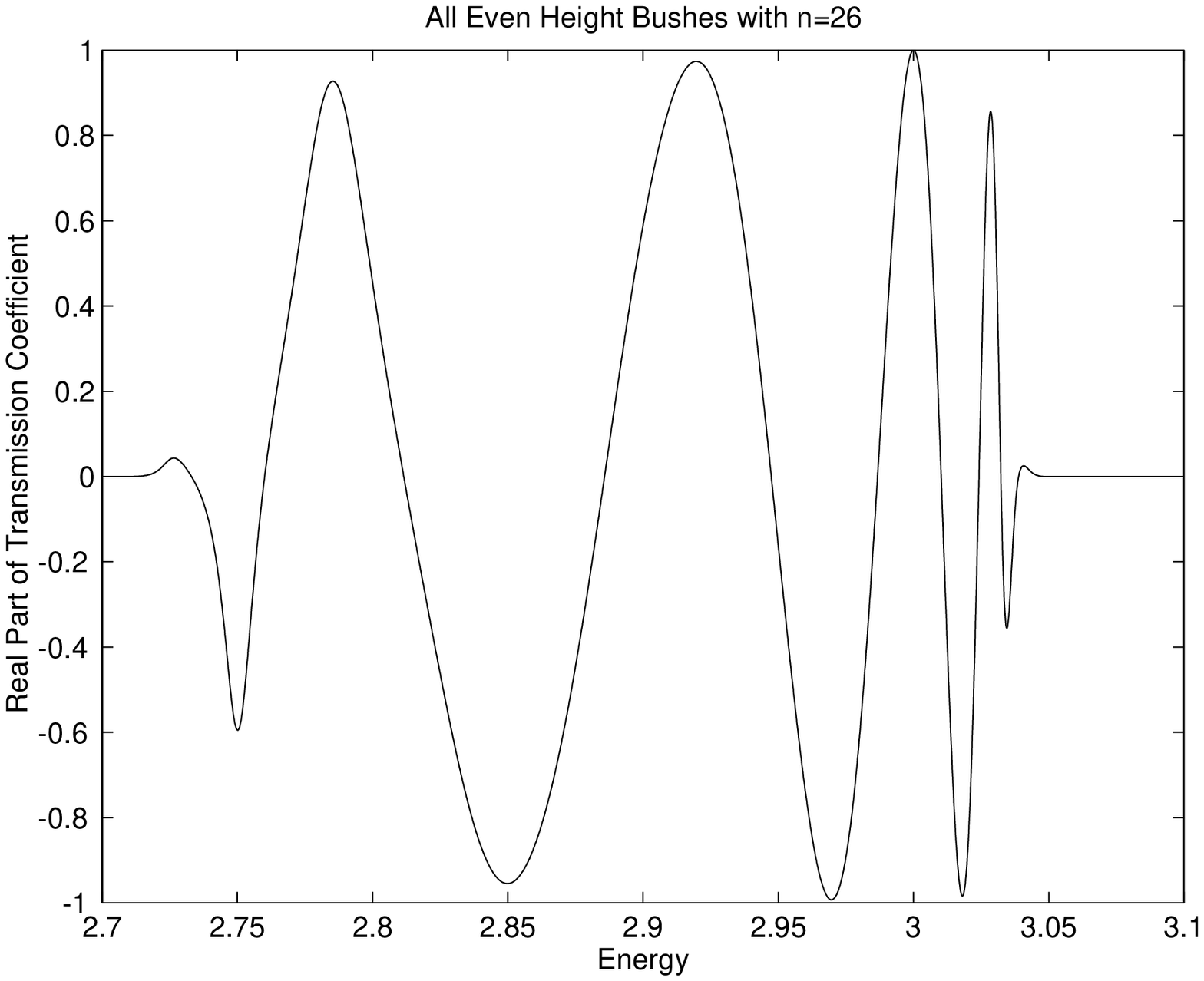}{900}{Figxi}{The real part of $T$ versus $E$ showing that $T$ does
not oscillate rapidly about zero close to where $T$ is~$1$, for the same tree as
Fig.\thinspace10.}%

\section{Implementing the Quantum System}

In this section, we show how to implement on a conventional quantum computer
the quantum systems previously described.  A conventional quantum computer
consists of $\ell$ spin $1/2$ particles that give rise to a $2^\ell$
dimensional complex Hilbert space with basis elements $|z_1z_2\cdots
z_\ell\rangle$ where we take each $z_i$ to be 0 or 1.  The computer program can
be thought of as a sequence of unitary operators $\hat{U}_\alpha$ each of which
acts on (at most)~$B$ bits.  That is, for each $\hat{U}_\alpha$
in the sequence, there is a set $S_\alpha = \{i_1,i_2,\dotsc,i_B\}$ that tells
us which
$B$ bits are being acted on and a $2^B$ by $2^B$ unitary matrix whose elements
we write as $U_\alpha (w_1'\cdots w_B';w_1\cdots w_B)$.  We then have for each
$\hat{U}_\alpha$,
\begin{equation}
\langle z_1'z_2'\cdots z_\ell ' |\hat{U}_\alpha|z_1z_2\cdots z_\ell\rangle \ = \
\prod_{j \notin S_\alpha} I(z_j=z'_j)U_\alpha(z'_{i_1}\cdots z'_{i_B};z_{i_1}\cdots
z_{i_B}) \ \ .
\label{eq:4.1}
\end{equation}
Here $I(s)$ is the indicator function that is 1 if $s$ is true and $0$ if $s$
is false.  This formula is just a way of writing that $\hat{U}_\alpha$
acts on $B$ bits.

In previous sections we described evolution through decision trees using the
quantum Hamiltonian $\hat{H}$ that gives rise to the unitary time evolution
operator $e^{-it\hat{H}}$.  To find a sequence of unitary operators each of
which acts on only several bits  and whose product gives (approximately) the
same evolution as $e^{-it\hat{H}}$, we follow the procedure given in
\cite{ref:3}.  Suppose 
\begin{equation}
\hat{H} = \sum^p_{k=1} \hat{H}_k
\label{eq:4.2}
\end{equation}
where for each $k$, $\hat{H}_k$ and hence $e^{-it\hat{H}_k}$ acts only on (at most) $B$
bits.  The Trotter formula says,
\begin{equation}
e^{-it\hat{H}}\approx \left[ e^{-it\hat{H}_1/m} \ e^{-it\hat{H}_2/m} \ \cdots \
e^{-it\hat{H}_p/m}\right]^m
\label{eq:4.3}
\end{equation}
for $t/m$ small.  Thus the evolution operator $e^{-it\hat{H}}$ can be
approximated as a product of $pm$ unitary operators each of which acts on a fixed
number of bits.  As a function of $n$ the largest times~$t$ that interest us are,
say, $n^A$. Taking
$m=n^{2A}$ allows us to obtain
$e^{-it\hat{H}}$ with a number of elementary unitary operators that only grows
polynomially with
$n$, as long as $p$ also grows only polynomially with $n$.

We now show two cases where the Hamiltonian $\hat{H}$ given by (\ref{eq:3}) can be
written as a sum of $\hat{H}_k$ where each $\hat{H}_k$ acts on a fixed number of
bits.  Consider first the underlying branching tree, \Figcite{Figi} and its associated
$\hat{H}$.  Start with $\ell = 2n+1$ bits that we group for convenience~as
\begin{equation}
(yx) = (y_0 y_1 \cdots y_n x_1\cdots x_n)\ \ .
\label{eq:4.4}
\end{equation}
The $y$ bits indicate the level of the node.  The states we use will have a
single $y_i =1$ and the rest $0$ to indicate that the node is at level $i$. 
The $x_1\cdots x_i$ will indicate the particular node at the $i^{\rm th}$ level;
these nodes will also have $x_{i+1} = x_{i+2} = \cdots = x_n = 0$.  We now define the
following one bit operators through their action on the basis vectors~$|yx
\rangle$:
\begin{eqnarray}
\hat{y}_j|yx \rangle &=& y_j |yx \rangle \nonumber \\
\hat{x}_j|yx \rangle &=& x_j |yx \rangle \nonumber \\
\hat{\rho}_j|yx \rangle &=& \hat{\rho}_j|y_0\cdots y_j \cdots y_nx \rangle = 
\bar{y}_j |y_0 \cdots \bar{y}_j \cdots y_nx \rangle 
\label{eq:4.5} \\
\hat{\sigma}_j|yx \rangle &=& \hat{\sigma}_j|yx_1 \cdots x_j\cdots x_n \rangle =
\bar{x}_j|y x_1 \cdots \bar{x}_j \cdots x_n\rangle   \nonumber
\end{eqnarray}
where $\bar{y}_j = 1-y_j$  and $\bar{x}_j = 1-x_j$.  We see that $\hat{x}_j$
and $\hat{y}_j$ are diagonal in this basis. 
The operator $\rdr$ acting on a state at level~$i$ brings it to level $i+1$ whereas
$\rrd$ moves from level $i+1$ to level~$i$.

The Hamiltonian (\ref{eq:3}) defined on the underlying branching tree is 
\begin{eqnarray}
\hat H &=& 2\hat y_0 + 3\sum_{i=1}^{n-1}\hat y_i + \hat y_n -\sum_{i=0}^{n-1}(\rdr +
\rrd)(1-\hat x_{i+1}) \nonumber\\
 &&\qquad {}-  \sum_{i=0}^{n-1}(\rdr \hat \sigma^{\phantom{\dagger}}_{i+1} +
 \rrd \hat \sigma^\dagger_{i+1})\ \ .
\label{eq:4.6}
\end{eqnarray}
The first three terms give the diagonal matrix elements. 
 The fourth term connects the
nodes $x_1\cdots x_i$ at level~$i$ with the nodes $x_1\cdots x_i0$ at level~$i+1$
whereas the last term connects $x_1\cdots x_i$ at level~$i$ with $x_1\cdots x_i1$ at
level~$i+1$. Thus we see that $\hat H$ can be written as a sum of $\hat H_k$ each of
which acts on at most three bits.

We have built a Hilbert space with $2^{2n+1}$ states whereas the underlying branching
tree has only $2^{n+1}-1$ nodes. However, if we start in the state corresponding to the
topmost node, that is, $y_0=1$ and all other bits~$0$, then if we act with $e^{-i\hat
Ht}$ with $\hat H$ given by (\ref{eq:4.6}) we only ever reach states in the subspace
corresponding to the underlying branching tree. The $2^{2n+1}$-dimensional Hilbert
space may not be the most economical choice to describe the tree but it suffices for our
purpose of showing that $\hat H$ can be built as a sum of local Hamiltonians.

Of course we also want to construct $\hat H$ as a sum of Hamiltonians acting on a fixed
number of bits for interesting trimmed decision trees. There are families of trimmed
trees whose Hamiltonians we cannot represent in this way. But for many interesting
problems we can write $\hat H$ as a sum of Hamiltonians that act on at most~$B$ bits,
where $B$ does not grow with~$n$. For example, we now show how to do this for a
version of the exact cover problem discussed in the introduction. We restrict the
matrix~$A$, which defines an instance of the exact cover problem, to have exactly
three~$1$'s in any row and three or fewer~$1$'s in any column. Even with this
restriction, the problem is NP-complete. 

Consider first the question of whether the $i^{\rm th}$-level node $x_1\cdots x_i$
connects to the $(i+1)^{\rm th}$-level node $x_1\cdots x_i1$. We assume that
$x_1\cdots x_i$ is in the tree and we need to be consistent with (\ref{eq:1}) so we
know that for each~$j$, $\sum_{k=1}^i A_{jk} x_k$ is~$0$ or~$1$. If for some~$j$ this
sum is~$1$ and also $A_{j,i+1}=1$, then $x_1\cdots x_i1$ is eliminated as a node.
Consider the function 
\begin{equation}
C_i^1 (x_1\cdots x_i) = \prod_{j=1}^m\Bigl\{
\bigl[ 1-\sum_{k=1}^i A_{jk} x_k \bigr]A_{j,i+1} + \bigl[1-A_{j,i+1}\bigr]\Bigr\}\ \ .
\label{eq:4.7}
\end{equation}
Given that $x_1\cdots x_i$ is an allowed node, then this function is~$1$ if $x_1\cdots
x_i1$ is allowed and~$0$ if $x_1\cdots x_i1$ is excluded. Furthermore, given the
restriction that $A$ has three~$1$'s in any row and three or fewer in any column, 
$C_i^1$ has at most six $x_k$'s appearing. 

Now we ask if $x_1\cdots x_i$ at level~$i$ connects to $x_1\cdots x_i0$ at level $i+1$.
This connection will be allowed unless for some~$j$ with $A_{j,i+1}=1$, there is a
$k\leq i$ and a distinct $k'\leq i$ such that $A_{jk}=A_{jk'}=1$ and $x_k=x_{k'}=0$. The
reason the node $x_1\cdots x_i0$ would be eliminated in this case is that there are
exactly three $1$'s in any row and~(\ref{eq:1}) could not be satisfied if the three bits
$x_k$,
$x_{k'}$, and $x_{i+1}$ are all~$0$. Now consider the function
\begin{equation}
d_i^j(x_1\cdots x_i) = \sum_{k=1}^i A_{jk}(1-x_k)\ \ .
\label{eq:4.8}
\end{equation}
For any $j$ with $A_{j,i+1}=1$, $d_i^j$ can be $0$, $1,$ or~$2$. Only if $d_i^j(x_1\cdots
x_i) = 2$ is $x_1\cdots x_i0$ eliminated. Let
\begin{equation}
C_i^0 (x_1\cdots x_i) = \prod_{j=1}^m\Bigl\{
\bigl[ \tfrac12 d_i^j(1-d_i^j)+1 \bigr]A_{j,i+1} + (1-A_{j,i+1})\Bigr\}\ \ .
\label{eq:4.9}
\end{equation}
Then this function is~$0$ if $x_1\cdots x_i0$ excluded and it is~$1$ if $x_1\cdots x_i0$ 
is allowed. Again because of the restrictions placed on~$A$, this function has only six
$x_k$'s appearing. 

The functions $C_i^0$ and $C_i^1$ can be promoted to operators simply by replacing
their arguments by the bit operators $\hat x_k$ defined in (\ref{eq:4.5}), that is, we
have $C_i^0 (\hat x_1\cdots \hat x_i)$ and $C_i^1 (\hat x_1\cdots \hat x_i)$. If we
multiply the last term in (\ref{eq:4.6}) by $C_i^1$ and the fourth term by $C_i^0$, the
Hamiltonian has off-diagonal elements only where the tree has connections. Similarly
we can write the diagonal term as
\begin{equation}
\hat H_{\rm diagonal} = 2\hat y_0 + \sum_{i=1}^{n-1} \hat y_i (1+ C_i^0 + C_i^1) +\hat
y_n\ \ .
\label{eq:4.10}
\end{equation}
Thus we have written the Hamiltonian for the trees trimmed by~$A$ in the
form~(\ref{eq:4.2}) with $B=9$. 

Generally, we think of decision trees as associated with functions~$f_i$ that impose
constraints: $f_i(x_1\cdots x_i)=1$ if the $(i-1)^{\rm th}$ level node $x_1\cdots
x_{i-1}$ is connected to the $i^{\rm th}$ level node $x_1\cdots x_i$; otherwise
$f_i=0$. The exact cover example above makes clear that as long as there is a fixed~$B$
such that $f_i(x_1\cdots x_i)$ depends on only~$B$ bits for each~$i$ ({\it which\/} bits
can vary with~$i$, of course) then the problem can be implemented within the usual
quantum computing paradigm -- we only need to replace $C_{i-1}^x (\hat x_1\cdots
\hat x_{i-1})$ in (\ref{eq:4.10}) by $f_i(\hat x_1\cdots \hat x_{i-1}, x)$ and also to
multiply the appropriate connection terms in (\ref{eq:4.6}) by $f_i(\hat x_1\cdots \hat
x_{i-1}, x)$.

Note that our example in Section~3 for which the quantum algorithm achieved
exponential speed-up does not meet this fixed-$B$ requirement. We do have, however,
similar examples that achieve exponential speed-up and that do meet this
requirement. These problems also rely on even-length, very structured bushes, and
also can be solved quickly by other classical algorithms.

\section{Conclusions}

There is great interest in devising quantum algorithms that improve on classical
algorithms, and there have been some notable successes. For example, the well-known
Shor~\cite{ref:6} and Grover~\cite{ref:5} algorithms demonstrate remarkable ingenuity.
Each uses quantum interference, the necessary ingredient for quantum speed-up, in
what appears to be a problem-specific way. So far these methods have not been
successfully applied to problems very different from the ones for which they were
originally devised. 

In this paper, we have considered a single time-dependent Hamiltonian that evolves a
quantum state through the nodes of a decision tree. 
(For a related approach, see \cite{ref:7}.)
This is in contrast to the usual
setup consisting of a sequence of unitary operators each acting on a fixed number of
bits. (For many problems, including NP-complete ones, our algorithm can be rewritten
in the conventional language of quantum computation.) Studying Hamiltonian evolution
on decision trees is facilitated by the technique of calculating energy-dependent
transmission coefficients. The example in Section Three shows explicitly how
interference allows a class of trees to be penetrated exponentially faster by quantum
evolution than by classical random walk. However, this example can be quickly solved
by a different classical algorithm. 

The particular Hamiltonian we chose allowed us to prove, in Section~Two, that the
quantum algorithm succeeded in polynomial time whenever the corresponding
classical random walk on the decision trees succeeded in polynomial time. In searching
for more examples where the quantum algorithm outperforms the classical algorithm,
one is not restricted to this Hamiltonian. We can imagine trying any Hamiltonian with
nonzero off-diagonal elements where there are links between the nodes on the
decision tree. With this flexibility, we hope that the class of trees that can be
penetrated quickly by the quantum algorithm is large enough to include classically
difficult problems. 

\subsubsection*{Acknowledgment}
We  thank Francis Low and Mike Sipser for their help and insights. We
also thank Rachel Cohen and Cindy Lewis for \LaTeX\ 
assistance, and Martin Stock for creating Figs.\thinspace\ref{fig:Figi}--\ref{fig:Figviii}
and final formatting. 
\vspace*{-1pc}

%% \begin{references} 


\begin{thebibliography}{99}
\bibitem{ref:1} ``Computational Complexity",  D.S.~Johnson and C.H.~Papadimitriou,
p.~37 of {\it The Traveling Salesman Problem}, E.L.~Lawler, J.K.~Lenstra, A.H.G.~Rinnooy
Kan, and D.B.~Shmoys, eds. John Wiley \& Sons, 1985.

\bibitem{ref:2}  A. Barenco, C.H. Bennett, R. Cleve, D.P.~DiVincenzo, N.~Margolus,
P.~Shor, T.~Sleator, J.A.~Smolin, and H.~Weinfurter,   Phys. Rev. {\bf A52}, 3457 (1995)
and references therein.

\bibitem{ref:3} S.~Lloyd, Science {\bf 273}, 1073 (1996).

\bibitem{ref:4} This is known as Gerschgorin's theorem; c.f.~C.G.~Cullen, {\it Matrices
and Linear Transformations}, $2^{\rm nd}$ edition, Addison-Wesley (1972), p.~283.

\bibitem{ref:5} L.K.~Grover, Proceedings, $28^{\rm th}$ Annual ACM Symposium
on the Theory of Computing (STOC) 1996, pp.~212--218 and quant-ph/9605043.

\bibitem{ref:6}  P.W.~Shor, ``Polynomial Time Algorithms for Prime Factorization and
Discrete Logarithms on a Quantum Computer", quant-ph/9508027

\bibitem{ref:7} T. Hogg, ``A Framework for Structured Quantum Search",
quant-ph/970113. 

%% \end{references}
\end{thebibliography}
\end{document}